\pdfoutput=1
      
\documentclass[lettersize,journal]{IEEEtran}

\usepackage{amsmath,amsfonts}
\usepackage{algorithmic}
\usepackage{algorithm}
\usepackage{array}
\usepackage{textcomp}
\usepackage{stfloats}
\usepackage{url}
\usepackage{verbatim}
\usepackage{cite}

\usepackage[acronyms,nonumberlist,nopostdot,nomain,nogroupskip,acronymlists={hidden}]{glossaries}
\usepackage{subcaption}
\usepackage[]{graphicx}
\usepackage{svg}
\usepackage{overpic}
\usepackage[section]{placeins}
\newsavebox{\measurebox}
\usepackage{multirow}
\usepackage{hhline}

\begin{document}
\newacronym{3gpp}{3GPP}{3rd Generation Partnership Project}
\newacronym{4g}{4G}{4th generation}
\newacronym{5g}{5G}{5th generation}
\newacronym{6g}{6G}{6th generation}
\newacronym{5gc}{5GC}{5G Core}
\newacronym{aau}{AAU}{Active Antenna Unit}
\newacronym{adc}{ADC}{Analog to Digital Converter}
\newacronym{aerpaw}{AERPAW}{Aerial Experimentation and Research Platform for Advanced Wireless}
\newacronym{ai}{AI}{Artificial Intelligence}
\newacronym{aimd}{AIMD}{Additive Increase Multiplicative Decrease}
\newacronym{am}{AM}{Acknowledged Mode}
\newacronym{amc}{AMC}{Adaptive Modulation and Coding}
\newacronym{amf}{AMF}{Access and Mobility Management Function}
\newacronym{aops}{AOPS}{Adaptive Order Prediction Scheduling}
\newacronym{api}{API}{Application Programming Interface}
\newacronym{apn}{APN}{Access Point Name}
\newacronym{ap}{AP}{Application Protocol}
\newacronym{aqm}{AQM}{Active Queue Management}
\newacronym{ausf}{AUSF}{Authentication Server Function}
\newacronym{avc}{AVC}{Advanced Video Coding}
\newacronym{awgn}{AGWN}{Additive White Gaussian Noise}
\newacronym{balia}{BALIA}{Balanced Link Adaptation Algorithm}
\newacronym{bbu}{BBU}{Base Band Unit}
\newacronym{bdp}{BDP}{Bandwidth-Delay Product}
\newacronym{ber}{BER}{Bit Error Rate}
\newacronym{bf}{BF}{Beamforming}
\newacronym{bler}{BLER}{Block Error Rate}
\newacronym{brr}{BRR}{Bayesian Ridge Regressor}
\newacronym{bs}{BS}{Base Station}
\newacronym{bsr}{BSR}{Buffer Status Report}
\newacronym{bss}{BSS}{Business Support System}
\newacronym{ca}{CA}{Carrier Aggregation}
\newacronym{caas}{CaaS}{Connectivity-as-a-Service}
\newacronym{cav}{CAV}{Connected and Autonoums Vehicle}
\newacronym{cb}{CB}{Code Block}
\newacronym{cc}{CC}{Congestion Control}
\newacronym{ccid}{CCID}{Congestion Control ID}
\newacronym{cco}{CC}{Carrier Component}
\newacronym{cd}{CD}{Continuous Delivery}
\newacronym{cdd}{CDD}{Cyclic Delay Diversity}
\newacronym{cdf}{CDF}{Cumulative Distribution Function}
\newacronym{cdn}{CDN}{Content Distribution Network}
\newacronym{cli}{CLI}{Command-line Interface}
\newacronym{cn}{CN}{Core Network}
\newacronym{codel}{CoDel}{Controlled Delay Management}
\newacronym{comac}{COMAC}{Converged Multi-Access and Core}
\newacronym{cord}{CORD}{Central Office Re-architected as a Datacenter}
\newacronym{cornet}{CORNET}{COgnitive Radio NETwork}
\newacronym{cosmos}{COSMOS}{Cloud Enhanced Open Software Defined Mobile Wireless Testbed for City-Scale Deployment}
\newacronym{cots}{COTS}{Commercial Off-the-Shelf}
\newacronym{cp}{CP}{Control Plane}
\newacronym{cpe}{CPE}{Customer Premises Equipment}
\newacronym{cyp}{CP}{Cyclic Prefix}
\newacronym{up}{UP}{User Plane}
\newacronym{cpu}{CPU}{Central Processing Unit}
\newacronym{cqi}{CQI}{Channel Quality Information}
\newacronym{cr}{CR}{Cognitive Radio}
\newacronym{cran}{CRAN}{Cloud \gls{ran}}
\newacronym{crs}{CRS}{Cell Reference Signal}
\newacronym{csi}{CSI}{Channel State Information}
\newacronym{csirs}{CSI-RS}{Channel State Information - Reference Signal}
\newacronym{cu}{CU}{Central Unit}
\newacronym{d2tcp}{D$^2$TCP}{Deadline-aware Data center TCP}
\newacronym{d3}{D$^3$}{Deadline-Driven Delivery}
\newacronym{dac}{DAC}{Digital to Analog Converter}
\newacronym{dag}{DAG}{Directed Acyclic Graph}
\newacronym{das}{DAS}{Distributed Antenna System}
\newacronym{dash}{DASH}{Dynamic Adaptive Streaming over HTTP}
\newacronym{dc}{DC}{Dual Connectivity}
\newacronym{dccp}{DCCP}{Datagram Congestion Control Protocol}
\newacronym{dce}{DCE}{Direct Code Execution}
\newacronym{dci}{DCI}{Downlink Control Information}
\newacronym{dctcp}{DCTCP}{Data Center TCP}
\newacronym{dl}{DL}{Downlink}
\newacronym{dmr}{DMR}{Deadline Miss Ratio}
\newacronym{dmrs}{DMRS}{DeModulation Reference Signal}
\newacronym{drlcc}{DRL-CC}{Deep Reinforcement Learning Congestion Control}
\newacronym{dsrc}{DSRC}
{dedicated short-range communications}
\newacronym{d2d}{D2D}{device-to-device}
\newacronym{drs}{DRS}{Discovery Reference Signal}
\newacronym{du}{DU}{Distributed Unit}
\newacronym{e2e}{E2E}{end-to-end}
\newacronym{earfcn}{EARFCN}{E-UTRA Absolute Radio Frequency Channel Number}
\newacronym{ecaas}{ECaaS}{Edge-Cloud-as-a-Service}
\newacronym{ecn}{ECN}{Explicit Congestion Notification}
\newacronym{edf}{EDF}{Earliest Deadline First}
\newacronym{embb}{eMBB}{Enhanced Mobile Broadband}
\newacronym{empower}{EMPOWER}{EMpowering transatlantic PlatfOrms for advanced WirEless Research}
\newacronym{enb}{eNB}{evolved Node Base}
\newacronym{endc}{EN-DC}{E-UTRAN-\gls{nr} \gls{dc}}
\newacronym{epc}{EPC}{Evolved Packet Core}
\newacronym{eps}{EPS}{Evolved Packet System}
\newacronym{es}{ES}{Edge Server}
\newacronym{etsi}{ETSI}{European Telecommunications Standards Institute}
\newacronym[firstplural=Estimated Times of Arrival (ETAs)]{eta}{ETA}{Estimated Time of Arrival}
\newacronym{eutran}{E-UTRAN}{Evolved Universal Terrestrial Access Network}
\newacronym{faas}{FaaS}{Function-as-a-Service}
\newacronym{fapi}{FAPI}{Functional Application Platform Interface}
\newacronym{fdd}{FDD}{Frequency Division Duplexing}
\newacronym{fdm}{FDM}{Frequency Division Multiplexing}
\newacronym{fdma}{FDMA}{Frequency Division Multiple Access}
\newacronym{fed4fire}{FED4FIRE+}{Federation 4 Future Internet Research and Experimentation Plus}
\newacronym{fir}{FIR}{Finite Impulse Response}
\newacronym{fit}{FIT}{Future \acrlong{iot}}
\newacronym{fpga}{FPGA}{Field Programmable Gate Array}
\newacronym{fr2}{FR2}{Frequency Range 2}
\newacronym{fr1}{FR1}{Frequency Range 1}
\newacronym{fs}{FS}{Fast Switching}
\newacronym{fscc}{FSCC}{Flow Sharing Congestion Control}
\newacronym{ftp}{FTP}{File Transfer Protocol}
\newacronym{fw}{FW}{Flow Window}
\newacronym{ge}{GE}{Gaussian Elimination}
\newacronym{gnb}{gNB}{Next Generation Node Base}
\newacronym{gop}{GOP}{Group of Pictures}
\newacronym{gpr}{GPR}{Gaussian Process Regressor}
\newacronym{gpu}{GPU}{Graphics Processing Unit}
\newacronym{gtp}{GTP}{GPRS Tunneling Protocol}
\newacronym{gtpc}{GTP-C}{GPRS Tunnelling Protocol Control Plane}
\newacronym{gtpu}{GTP-U}{GPRS Tunnelling Protocol User Plane}
\newacronym{gtpv2c}{GTPv2-C}{\gls{gtp} v2 - Control}
\newacronym{gw}{GW}{Gateway}
\newacronym{harq}{HARQ}{Hybrid Automatic Repeat reQuest}
\newacronym{hetnet}{HetNet}{Heterogeneous Network}
\newacronym{hh}{HH}{Hard Handover}
\newacronym{hol}{HOL}{Head-of-Line}
\newacronym{hqf}{HQF}{Highest-quality-first}
\newacronym{hss}{HSS}{Home Subscription Server}
\newacronym{http}{HTTP}{HyperText Transfer Protocol}
\newacronym{ia}{IA}{Initial Access}
\newacronym{iab}{IAB}{Integrated Access and Backhaul}
\newacronym{ic}{IC}{Incident Command}
\newacronym{ietf}{IETF}{Internet Engineering Task Force}
\newacronym{ieee}{IEEE}{Institute of Electrical and Electronics Engineers}
\newacronym{imsi}{IMSI}{International Mobile Subscriber Identity}
\newacronym{iot}{IoT}{Internet of Things}
\newacronym{ip}{IP}{Internet Protocol}
\newacronym{itu}{ITU}{International Telecommunication Union}
\newacronym{kpi}{KPI}{Key Performance Indicator}
\newacronym{kpm}{KPM}{Key Performance Measurement}
\newacronym{kvm}{KVM}{Kernel-based Virtual Machine}
\newacronym{los}{LoS}{Line of Sight}
\newacronym{lsm}{LSM}{Link-to-System Mapping}
\newacronym{lstm}{LSTM}{Long Short Term Memory}
\newacronym{lte}{LTE}{Long Term Evolution}
\newacronym{lxc}{LXC}{Linux Container}
\newacronym{m2m}{M2M}{Machine to Machine}
\newacronym{mac}{MAC}{Medium Access Control}
\newacronym{manet}{MANET}{Mobile Ad Hoc Network}
\newacronym{mano}{MANO}{Management and Orchestration}
\newacronym{mc}{MC}{Multi-Connectivity}
\newacronym{mcc}{MCC}{Mobile Cloud Computing}
\newacronym{mchem}{MCHEM}{Massive Channel Emulator}
\newacronym{mcs}{MCS}{Modulation and Coding Scheme}
\newacronym{mec2}{MEC}{Multi-access Edge Computing}
\newacronym{mec}{MEC}{Mobile Edge Computing}
\newacronym{mfc}{MFC}{Mobile Fog Computing}
\newacronym{mgen}{MGEN}{Multi-Generator}
\newacronym{mi}{MI}{Mutual Information}
\newacronym{mib}{MIB}{Master Information Block}
\newacronym{miesm}{MIESM}{Mutual Information Based Effective SINR}
\newacronym{mimo}{MIMO}{Multiple Input, Multiple Output}
\newacronym{ml}{ML}{Machine Learning}
\newacronym{mlr}{MLR}{Maximum-local-rate}
\newacronym[plural=\gls{mme}s,firstplural=Mobility Management Entities (MMEs)]{mme}{MME}{Mobility Management Entity}
\newacronym{mmtc}{mMTC}{Massive Machine-Type Communications}
\newacronym{mmwave}{mmWave}{millimeter wave}
\newacronym{mno}{MNO}{Mobile Network Operators}
\newacronym{mpdccp}{MP-DCCP}{Multipath Datagram Congestion Control Protocol}
\newacronym{mptcp}{MPTCP}{Multipath TCP}
\newacronym{mr}{MR}{Maximum Rate}
\newacronym{mrdc}{MR-DC}{Multi \gls{rat} \gls{dc}}
\newacronym{mse}{MSE}{Mean Square Error}
\newacronym{mss}{MSS}{Maximum Segment Size}
\newacronym{mt}{MT}{Mobile Termination}
\newacronym{mtd}{MTD}{Machine-Type Device}
\newacronym{mtu}{MTU}{Maximum Transmission Unit}
\newacronym{mumimo}{MU-MIMO}{Multi-user \gls{mimo}}
\newacronym{mvno}{MVNO}{Mobile Virtual Network Operator}
\newacronym{nalu}{NALU}{Network Abstraction Layer Unit}
\newacronym{nas}{NAS}{Network Attached Storage}
\newacronym{nat}{NAT}{Network Address Translation}
\newacronym{nbiot}{NB-IoT}{Narrow Band IoT}
\newacronym{nfv}{NFV}{Network Function Virtualization}
\newacronym{nfvi}{NFVI}{Network Function Virtualization Infrastructure}
\newacronym{ni}{NI}{Network Interfaces}
\newacronym{nic}{NIC}{Network Interface Card}
\newacronym{now}{NOW}{Non Overlapping Window}
\newacronym{nsm}{NSM}{Network Service Mesh}
\newacronym{nr}{NR}{New Radio}
\newacronym{nrf}{NRF}{Network Repository Function}
\newacronym{nr-u}{NR-U}{New Radio Unlicensed}
\newacronym{nsa}{NSA}{Non Stand Alone}
\newacronym{nse}{NSE}{Network Slicing Engine}
\newacronym{nssf}{NSSF}{Network Slice Selection Function}
\newacronym{oai}{OAI}{OpenAirInterface}
\newacronym{oaicn}{OAI-CN}{\gls{oai} \acrlong{cn}}
\newacronym{oairan}{OAI-RAN}{\acrlong{oai} \acrlong{ran}}
\newacronym{oam}{OAM}{Operations, Administration and Maintenance}
\newacronym{ofdm}{OFDM}{Orthogonal Frequency Division Multiplexing}
\newacronym{olia}{OLIA}{Opportunistic Linked Increase Algorithm}
\newacronym{omec}{OMEC}{Open Mobile Evolved Core}
\newacronym{onap}{ONAP}{Open Network Automation Platform}
\newacronym{onf}{ONF}{Open Networking Foundation}
\newacronym{onos}{ONOS}{Open Networking Operating System}
\newacronym{oom}{OOM}{\gls{onap} Operations Manager}
\newacronym{opnfv}{OPNFV}{Open Platform for \gls{nfv}}
\newacronym{oran}{O-RAN}{Open \gls{ran}}
\newacronym{orbit}{ORBIT}{Open-Access Research Testbed for Next-Generation Wireless Networks}
\newacronym{os}{OS}{Operating System}
\newacronym{oss}{OSS}{Operations Support System}
\newacronym{pa}{PA}{Position-aware}
\newacronym{pase}{PASE}{Prioritization, Arbitration, and Self-adjusting Endpoints}
\newacronym{pawr}{PAWR}{Platforms for Advanced Wireless Research}
\newacronym{pbch}{PBCH}{Physical Broadcast Channel}
\newacronym{pcef}{PCEF}{Policy and Charging Enforcement Function}
\newacronym{pcfich}{PCFICH}{Physical Control Format Indicator Channel}
\newacronym{pcrf}{PCRF}{Policy and Charging Rules Function}
\newacronym{pdcch}{PDCCH}{Physical Downlink Control Channel}
\newacronym{pdcp}{PDCP}{Packet Data Convergence Protocol}
\newacronym{pdsch}{PDSCH}{Physical Downlink Shared Channel}
\newacronym{pdu}{PDU}{Packet Data Unit}
\newacronym{pf}{PF}{Proportional Fair}
\newacronym{pgw}{PGW}{Packet Gateway}
\newacronym{phich}{PHICH}{Physical Hybrid ARQ Indicator Channel}
\newacronym{phy}{PHY}{Physical}
\newacronym{pmch}{PMCH}{Physical Multicast Channel}
\newacronym{pmi}{PMI}{Precoding Matrix Indicators}
\newacronym{powder}{POWDER}{Platform for Open Wireless Data-driven Experimental Research}
\newacronym{ppo}{PPO}{Proximal Policy Optimization}
\newacronym{ppp}{PPP}{Poisson Point Process}
\newacronym{prach}{PRACH}{Physical Random Access Channel}
\newacronym{prb}{PRB}{Physical Resource Block}
\newacronym{psnr}{PSNR}{Peak Signal to Noise Ratio}
\newacronym{pss}{PSS}{Primary Synchronization Signal}
\newacronym{pucch}{PUCCH}{Physical Uplink Control Channel}
\newacronym{pusch}{PUSCH}{Physical Uplink Shared Channel}
\newacronym{qam}{QAM}{Quadrature Amplitude Modulation}
\newacronym{qci}{QCI}{\gls{qos} Class Identifier}
\newacronym{qoe}{QoE}{Quality of Experience}
\newacronym{qos}{QoS}{Quality of Service}
\newacronym{quic}{QUIC}{Quick UDP Internet Connections}
\newacronym{ra}{RA}{Resouces Allocation}
\newacronym{rach}{RACH}{Random Access Channel}
\newacronym{ran}{RAN}{Radio Access Network}
\newacronym[firstplural=Radio Access Technologies (RATs)]{rat}{RAT}{Radio Access Technology}
\newacronym{rbg}{RBG}{Resource Block Group}
\newacronym{rcn}{RCN}{Research Coordination Network}
\newacronym{rc}{RC}{RAN Control}
\newacronym{rec}{REC}{Radio Edge Cloud}
\newacronym{red}{RED}{Random Early Detection}
\newacronym{renew}{RENEW}{Reconfigurable Eco-system for Next-generation End-to-end Wireless}
\newacronym{rf}{RF}{Radio Frequency}
\newacronym{rfc}{RFC}{Request for Comments}
\newacronym{rfr}{RFR}{Random Forest Regressor}
\newacronym{ric}{RIC}{\gls{ran} Intelligent Controller}
\newacronym{rlc}{RLC}{Radio Link Control}
\newacronym{rlf}{RLF}{Radio Link Failure}
\newacronym{rlnc}{RLNC}{Random Linear Network Coding}
\newacronym{rmr}{RMR}{RIC Message Router}
\newacronym{rmse}{RMSE}{Root Mean Squared Error}
\newacronym{rnis}{RNIS}{Radio Network Information Service}
\newacronym{rr}{RR}{Round Robin}
\newacronym{rrc}{RRC}{Radio Resource Control}
\newacronym{rrm}{RRM}{Radio Resource Management}
\newacronym{rru}{RRU}{Remote Radio Unit}
\newacronym{rs}{RS}{Remote Server}
\newacronym{rsrp}{RSRP}{Reference Signal Received Power}
\newacronym{rsrq}{RSRQ}{Reference Signal Received Quality}
\newacronym{rss}{RSS}{Received Signal Strength}
\newacronym{rssi}{RSSI}{Received Signal Strength Indicator}
\newacronym{rtt}{RTT}{Round Trip Time}
\newacronym{ru}{RU}{Radio Unit}
\newacronym{rus}{RSU}{Road Side Unit}
\newacronym{rw}{RW}{Receive Window}
\newacronym{rx}{RX}{Receiver}
\newacronym{s1ap}{S1AP}{S1 Application Protocol}
\newacronym{sa}{SA}{standalone}
\newacronym{sack}{SACK}{Selective Acknowledgment}
\newacronym{sap}{SAP}{Service Access Point}
\newacronym{sc2}{SC2}{Spectrum Collaboration Challenge}
\newacronym{scef}{SCEF}{Service Capability Exposure Function}
\newacronym{sch}{SCH}{Secondary Cell Handover}
\newacronym{scs}{SCS}{Sub-Carrier Spacing}
\newacronym{scoot}{SCOOT}{Split Cycle Offset Optimization Technique}
\newacronym{sctp}{SCTP}{Stream Control Transmission Protocol}
\newacronym{sdap}{SDAP}{Service Data Adaptation Protocol}
\newacronym{sdk}{SDK}{Software Development Kit}
\newacronym{sdm}{SDM}{Space Division Multiplexing}
\newacronym{sdma}{SDMA}{Spatial Division Multiple Access}
\newacronym{sdn}{SDN}{Software-defined Networking}
\newacronym{sdr}{SDR}{Software-defined Radio}
\newacronym{seba}{SEBA}{SDN-Enabled Broadband Access}
\newacronym{sgsn}{SGSN}{Serving GPRS Support Node}
\newacronym{sgw}{SGW}{Service Gateway}
\newacronym{si}{SI}{Study Item}
\newacronym{sib}{SIB}{Secondary Information Block}
\newacronym{sinr}{SINR}{Signal to Interference plus Noise Ratio}
\newacronym{sip}{SIP}{Session Initiation Protocol}
\newacronym{siso}{SISO}{Single Input, Single Output}
\newacronym{sla}{SLA}{Service Level Agreement}
\newacronym{sm}{SM}{Service Model}
\newacronym{smo}{SMO}{Service Management and Orchestration}
\newacronym{smsgmsc}{SMS-GMSC}{\gls{sms}-Gateway}
\newacronym{snr}{SNR}{Signal-to-Noise-Ratio}
\newacronym{son}{SON}{Self-Organizing Network}
\newacronym{sptcp}{SPTCP}{Single Path TCP}
\newacronym{srb}{SRB}{Service Radio Bearer}
\newacronym{srn}{SRN}{Standard Radio Node}
\newacronym{srs}{SRS}{Sounding Reference Signal}
\newacronym{ss}{SS}{Synchronization Signal}
\newacronym{sss}{SSS}{Secondary Synchronization Signal}
\newacronym{st}{ST}{Spanning Tree}
\newacronym{svc}{SVC}{Scalable Video Coding}
\newacronym{tb}{TB}{Transport Block}
\newacronym{tcp}{TCP}{Transmission Control Protocol}
\newacronym{tdd}{TDD}{Time Division Duplexing}
\newacronym{tdm}{TDM}{Time Division Multiplexing}
\newacronym{tdma}{TDMA}{Time Division Multiple Access}
\newacronym{tfl}{TfL}{Transport for London}
\newacronym{tfrc}{TFRC}{TCP-Friendly Rate Control}
\newacronym{tft}{TFT}{Traffic Flow Template}
\newacronym{tgen}{TGEN}{Traffic Generator}
\newacronym{tip}{TIP}{Telecom Infra Project}
\newacronym{tm}{TM}{Transparent Mode}
\newacronym{to}{TO}{Telco Operator}
\newacronym{tr}{TR}{Technical Report}
\newacronym{trp}{TRP}{Transmitter Receiver Pair}
\newacronym{ts}{TS}{Technical Specification}
\newacronym{tti}{TTI}{Transmission Time Interval}
\newacronym{ttt}{TTT}{Time-to-Trigger}
\newacronym{tue}{TUE}{Test UE}
\newacronym{tx}{TX}{Transmitter}
\newacronym{u6g}{U6G}{Upper 6GHz}
\newacronym{uas}{UAS}{Unmanned Aerial System}
\newacronym{uav}{UAV}{Unmanned Aerial Vehicle}
\newacronym{udm}{UDM}{Unified Data Management}
\newacronym{udp}{UDP}{User Datagram Protocol}
\newacronym{udr}{UDR}{Unified Data Repository}
\newacronym{ue}{UE}{User Equipment}
\newacronym{uhd}{UHD}{\gls{usrp} Hardware Driver}
\newacronym{ul}{UL}{Uplink}
\newacronym{um}{UM}{Unacknowledged Mode}
\newacronym{uml}{UML}{Unified Modeling Language}
\newacronym{upa}{UPA}{Uniform Planar Array}
\newacronym{upf}{UPF}{User Plane Function}
\newacronym{urllc}{URLLC}{Ultra Reliable and Low Latency Communications}
\newacronym{usa}{U.S.}{United States}
\newacronym{usim}{USIM}{Universal Subscriber Identity Module}
\newacronym{usrp}{USRP}{Universal Software Radio Peripheral}
\newacronym{utc}{UTC}{Urban Traffic Control}
\newacronym{vim}{VIM}{Virtualization Infrastructure Manager}
\newacronym{vm}{VM}{Virtual Machine}
\newacronym{vnf}{VNF}{Virtual Network Function}
\newacronym{volte}{VoLTE}{Voice over \gls{lte}}
\newacronym{voltha}{VOLTHA}{Virtual OLT HArdware Abstraction}
\newacronym{vr}{VR}{Virtual Reality}
\newacronym{vran}{vRAN}{Virtualized \gls{ran}}
\newacronym{vss}{VSS}{Video Streaming Server}
\newacronym{v2x}{V2X}{vehicle-to-everything}
\newacronym{v2i}{V2I}{vehicle-to-infrastructure}
\newacronym{v2v}{V2V}{vehicle-to-vehicle}
\newacronym{v2n}{V2N}{vehicle-to-network}
\newacronym{wbf}{WBF}{Wired Bias Function}
\newacronym{wf}{WF}{Waterfilling}
\newacronym{wg}{WG}{Working Group}
\newacronym{wlan}{WLAN}{Wireless Local Area Network}
\newacronym{wrc}{WRC}{World Radiocommunication Conference}
\newacronym{osm}{OSM}{Open Source \gls{nfv} Management and Orchestration}
\newacronym{pnf}{PNF}{Physical Network Function}
\newacronym{drl}{DRL}{Deep Reinforcement Learning}
\newacronym{mtc}{MTC}{Machine-type Communications}
\newacronym{osc}{OSC}{O-RAN Software Community}
\newacronym{mns}{MnS}{Management Services}
\newacronym{ves}{VES}{\gls{vnf} Event Stream}
\newacronym{ei}{EI}{Enrichment Information}
\newacronym{fh}{FH}{Fronthaul}
\newacronym{fft}{FFT}{Fast Fourier Transform}
\newacronym{laa}{LAA}{Licensed-Assisted Access}
\newacronym{plfs}{PLFS}{Physical Layer Frequency Signals}
\newacronym{ptp}{PTP}{Precision Time Protocol}
\newacronym{lidar}{LiDAR}{Light Detection And Ranging}
\newacronym{dem}{DEM}{Digital Elevation Model}
\newacronym{dtm}{DEM}{Digital Terrain Model}
\newacronym{dsm}{DEM}{Digital Surface Models}
\newacronym{ota}{OTA}{Over-The-Air}
\newacronym{ns}{NS}{Network Slicing}
\newacronym{ne}{NE}{Nash Equilibrium}
\newacronym{hf}{HF}{High Frequency}
\newacronym{noma}{NOMA}{Non-Orthogonal Multiple Access}
\newacronym{sre}{SRE}{Smart Radio Environment}
\newacronym{ris}{RIS}{Reconfigurable Intelligent Surface}
\newacronym{inp}{InP}{Infrastructure Provider}
\newacronym{smf}{SMF}{Slicing Magangement Framework}
\newacronym{nsn}{NSN}{Network Slicing Negotiation}
\newacronym{sms}{SMS}{Slicing MAC Scheduler}
\newacronym{brd}{BRD}{Best Response Dynamics}
\newacronym{dssbr}{DSSBR}{Double Step Smoothed Best Response}
\newacronym{poa}{PoA}{Price of Anarchy}
\newacronym{pos}{PoS}{Price of Stability}
\newacronym{milp}{MILP}{Mixed Integer-Linear Program}
\newacronym{pod}{PoD}{Price of DSSBR}
\newacronym{roc}{ROC}{Radio Overload Control}
\newacronym{ciot}{cIoT}{critical Internet of Things}
\newacronym{embbpr}{eMBB Pr.}{enhanced Mobile BroadBand Premium}
\newacronym{sps}{SPS}{Semi-persistent Scheduling}
\newacronym{cg}{CG}{Configured Grant}
\newacronym{embbbs}{eMBB Bs.}{enhanced Mobile BroadBand Basic}
\newacronym{en}{EN}{Edge Node}
\newacronym{ec}{EC}{Edge Computing}
\newacronym{sp}{SP}{Service Provider}
\newacronym{me}{ME}{Market Equilibrium}
\newacronym{so}{SO}{Social Optimum}
\newacronym{wso}{WSO}{Weighted Social Optimum}
\newacronym{ps}{PS}{Proportional Sharing}
\newacronym{eg}{EG}{Eisenberg-Gale program}
\newacronym{pe}{PE}{Pareto Efficiency}
\newacronym{nsw}{NSW}{Nash Social Welfare}
\newacronym{ef}{EF}{Envy-Freeness}
\newacronym{sub6}{sub6GHz}{Below 6GHz}
\newacronym{ncr}{NCR}{Network-Controlled Repeater}
\newacronym{nlos}{NLoS}{Non-Line of Sight}
\newacronym{src}{SRC}{Smart Radio Connection}
\newacronym{srd}{SRD}{Smart Radio Device}
\newacronym{cs}{CS}{Candidate Site}
\newacronym{tp}{TP}{Test Point}
\newacronym{fov}{FoV}{Field of View}
\newacronym{nrric}{near-RT RIC}{Near Real-time {RAN} Intelligent Controller}
\newacronym{e2ap}{E2AP}{E2 Application Protocol}
\newacronym{e2sm}{E2SM}{E2 Service Model}
\newacronym{nrtric}{non-RT RIC}{Non-Real-Time {RIC}}
\newacronym{itti}{ITTI}{Inter-task Interface}
\newacronym{bap}{BAP}{Backhaul Adaptation Protocol}
\newacronym{iabest}{IABEST}{Integrated Access and Backhaul Experimental large-Scale Tetbed}
\newacronym{teid}{TEID}{Tunnel Endpoint Identifier}
\newacronym{dlsch}{DL-SCH}{Downlink Shared Channel }
\newacronym{ulsch}{UL-SCH}{Uplink Shared Channel }
\newacronym{rsu}{RSU}{Road Side Unit}
\newacronym{its}{ITS}{Intelligent Transportation Systems}
\newacronym{vanet}{VANET}{Vehicular Ad-hoc Network}
\newacronym{dt}{DT}{Digital Twin}
\newacronym{ecc}{ECC}{Edge Computing Cluster}
\newacronym{o2i}{O2I}{Outdoor-to-indoor}
\newacronym{fwa}{FWA}{Fixed Wireless Access}
\newacronym{afc}{AFC}{Automated Frequency Coordinator}

\title{Exploring Upper-6GHz and mmWave in Real-World 5G Networks: A Direct on-Field Comparison}

\author{Marcello Morini, Eugenio Moro, Ilario Filippini, Antonio Capone, Danilo De Donno\thanks{M. Morini, E. Moro, I. Filippini and A. Capone were with DEIB, Politecnico di Milano, Milan,
Italy. \textit{\{name.surname\}@polimi.it}}
\thanks{D. De Donno was with Milan Research Center, Huawei Technologies Italia S.r.l., Milan, Italy. \textit{danilo.dedonno@huawei.com}}%
}

\markboth{Journal of \LaTeX\ Class Files,~Vol.~14, No.~8, August~2021}%
{Shell \MakeLowercase{\textit{et al.}}: A Sample Article Using IEEEtran.cls for IEEE Journals}

\maketitle

\newcommand{\ed}{\color{red}}
\begin{abstract}
    %
%
%
The spectrum crunch challenge poses a vital threat to the progress of cellular networks and recently prompted the inclusion of \gls{mmwave} and \gls{u6g} in the 3GPP standards. These two bands promise to unlock a large portion of untapped spectrum, but the harsh propagation due to the increased carrier frequency might negatively impact the performance of urban \gls{ran} deployments. Within the span of a year, two co-located 5G networks operating in these frequency bands were deployed at Politecnico di Milano, Milan, Italy, entirely dedicated to the dense urban performance assessment of the two systems. This paper presents an in-depth analysis of the measurement campaigns conducted on them, with the \gls{u6g} campaign representing the first of its kind. A benchmark is provided by ray-tracing simulations. The results suggest that networks operating in these frequency bands provide good indoor and outdoor coverage and throughput in urban scenarios, even when deployed in the macro base station setup common to lower frequencies. In addition, a comparative performance analysis of these two key technologies is provided, offering insights on their relative strengths, weaknesses and improvement margins and informing on which bands is better suited for urban macro coverage.
    \label{sec:abstract}
\end{abstract}

\begin{IEEEkeywords} 
5G, measurements, millimeter-wave, upper-6GHz, commercial deployment, Milan.
\end{IEEEkeywords}

\section{Introduction}
\label{sec:intro}
The ongoing rise in the number of mobile users and their requirements for bandwidth makes the exploration of new spectrum necessary~\cite{rappaport2013millimeter}. The overcrowding of lower frequency bands, where the propagation is good and the technology is familiar, led to what is known as \textit{spectrum crunch}, which jeopardizes the future performances of mobile radio networks \cite{spectrum_crunch}.
Two bands in particular attracted attention as a prompt way out to mitigate such shortage in the next years, namely the 6-GHz and the \gls{mmwave} bands.

The frequency range from $5.925$ to $7.125\,$GHz, known as the $6\,$GHz band, possesses good coverage features that characterize the mid-band spectrum. It provides sufficient bandwidths to ensure high-speed data transfer --- letting users reach gigabits per second throughput --- without the need to resort to higher frequencies. These qualities led to the introduction of the \acrfull{u6g} among \gls{nr} bands in \gls{3gpp} Release 17 and, in parallel, to the birth of Wi-Fi 6E within the Wi-Fi Alliance.\\ 
However, the different and incompatible types of spectrum access that these two standards require (i.e., licensed and unlicensed, respectively), brought the 6-GHz licensing process on a winding path. 
In April 2020, the United States Federal Communications Commission (FCC) enabled the use of the entire 1.200 MHz spectrum for unlicensed use~\cite{FCC-20-51}, particularly for low-power indoor applications under an Automatic Frequency Coordination (AFC) framework. 
In China, the Ministry of Industry and Information Technology started supporting licensing policies at the end of June 2023, when it officially endorsed the \gls{u6g} (or portions thereof) for licensed systems\footnote{\url{https://www.miit.gov.cn/jgsj/wgj/gzdt/art/2023/art_92c8962a03a44a37becc2963cb3c8df9.html}}. 
For Europe, Africa, and part of Asia, the discussion had a turning point in December 2023, at the World Radiocommunication Conference 2023. After years of technical analysis and discussions, the International Telecommunication Union decided to split the 6-GHz band into lower (5.945–6.425 GHz) and upper part (6.425–7.125 GHz) and to allocate only the latter for licensed use \cite{wrc_decision}. Europe is expected to adopt this decision according to the Radio Spectrum Policy Group (i.e., the high advisory body of the European Commission) opinion on \gls{wrc} \cite{rspg22}.

\Gls{mmwave} represents another asset for future mobile networks. Firstly embedded in mobile access networks in 2012 to support high throughputs in Wi-Fi WiGig networks, \gls{mmwave} were then brought into 5G starting in 2017 with Release 15. Frequencies between 24.25 GHz and 71 GHz are currently supported by 3GPP 5G standards as well as IEEE 802.11ad, aj, and ay.
According to 5G standards, the total bandwidth in \gls{fr2} is around 29 GHz, which is six times more than that available in \gls{fr1}~\cite{fr2, fr1}. 
This free spectrum real estate is meant to be one of the key assets to reach the goals of IMT-2020 \cite{imt2020_requirements}. Its role is vital for \gls{embb} application, in particular in hotspots and dense urban scenarios. However, its adoption has not reached the expected scale so far. To December 2023, only a minority of the $146000$ 5G base stations\footnote{\url{https://www.speedtest.net/it/ookla-5g-map}} deployed around the world use \gls{fr2} \cite{mmwave_development_2023}. \\
High-frequency bands in general, suffer from harsher propagation than low and mid bands. However, when put into the context of mobile radio networks, some encouraging signals should be considered. Free-space path loss increases with frequency, but smaller wavelengths facilitate integrating very directive \gls{mimo} antennas, eventually generating higher gain and narrow beams that offset the path loss. Atmospheric gaseous absorptions oscillate from a minimum of 0.1 dB/km at 35HGz to a maximum of 10 dB/km at 60GHz due to a peak in oxygen absorption \cite{mmwave_propagation}. Still, these values should not be too alarming, considering that today's cell radius in urban environments is around 200m \cite{rappaport2013millimeter}. The same reasoning can be extended for rain attenuation: while it is true that at 26 GHz, heavy rain (25mm/h) attenuates around 4dB/km \cite{mmwave_propagation}, this value does not heavily affect the connection on cell-sized distances. Certainly, the power originated by diffraction is smaller at \gls{mmwave} than at lower frequencies. This per-se limits \gls{nlos} propagation. However, as shown in the following, the effect might not be so severe in urban environments.
Outdoor-to-indoor propagation is perhaps the most severe scenario, and high throughput connections can only be established in the presence of clear, thin glass \cite{Outdoor_to_Indoor, rappaport2013millimeter}, as it will be shown in this work. 

We believe that, as in any engineering discussion, there is a need for experiments and results to understand the benefits and challenges of the alternatives under investigation. 
Pursuing this objective, we deployed two standard-compliant 5G networks working in the \gls{u6g} and \gls{mmwave} bands, both configured as macro base stations. The co-location of the two deployments at the exact same point in the city of Milan, Italy, also allows a direct and meaningful comparison.
Both networks present distinctive features. At the time of writing (Dec. 23), no other contributions on U6G 5G deployments are present, making this the first measurement campaign available in the literature at such frequencies \cite{will_U6G_work}. Furthermore, our \gls{mmwave} network stands out for not relying on anchoring base stations at lower frequencies (i.e., fallback to sub-6GHz connection), thus providing a clear and unbiased perspective on this band's performances. Finally, the comparative analysis enabled by the co-location of the two systems reveals further insights on the ideal application scenarios and the performance improvement margins. 

The remainder of this paper is organized as follows.\textit{ Section \ref{sec:scenario}} introduces the measurement scenario and characterizes it with ray-tracing simulations.\textit{ Section \ref{sec:6GHz}} and \textit{\ref{sec:mmWave}} present the results of the U6G and mmWave measurement campaigns, respectively, which are then compared in \textit{Section \ref{sec:comparison}}. The related works are reported in \textit{Section \ref{sec:related}}, just before concluding the paper in \textit{Section \ref{sec:conclusion}}.

\section{Measurement scenarios and equipment}
\label{sec:scenario}
\begin{figure*}[t!]
    \centering
    \includegraphics[width=2\columnwidth]{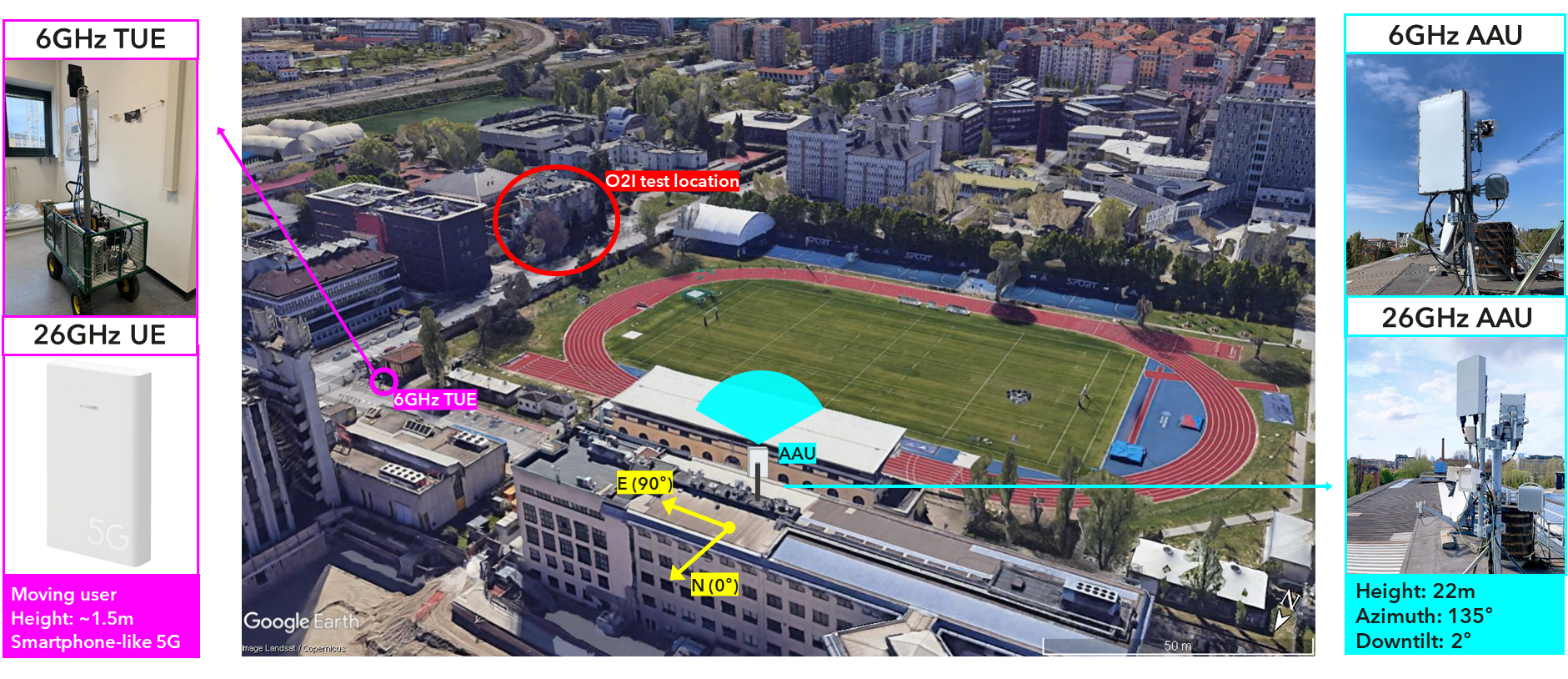}
    \caption{Aerial view of the test area with equipment details}  
    \label{fig:aerial_view_with_equipments}
\end{figure*}

The Upper 6GHz and millimeter-wave networks were mounted on the rooftop of one building in the main campus of Politecnico di Milano, in Milan, Italy. In this section, the common aspects between the deployments are stated, followed by the ray-tracing simulation results. Deployments' specific features are then outlined and the measurement methodology is eventually reported.

Both networks are made of three main components: an \gls{aau}, a \gls{bbu} and a 5G \gls{cn} deployment.
The \gls{aau} was installed at a height of 22 meters, down-tilted by 2°, and covering a 120° sector with a center azimuth of 135°.
The spot and the orientation were chosen exactly equal to enable a meaningful comparison.
The selected site location, along with the main hardware pieces composing the testbed, are depicted in \textit{Figure \ref{fig:aerial_view_with_equipments}}. Equipment details are reported in \textit{Tab. \ref{tab:hardware_details}} and commented in the remaining of this section. \\
In both deployments, the \glspl{bbu} and \glspl{aau} had been connected by a 25Gbps fiber fronthaul and mounted on the rooftop. The \glspl{bbu} was then connected to the \gls{cn} using a 10 Gbps backhaul link. \\
On the user side, the two campaigns leveraged different equipment, which are described below. Each \gls{ue} was mounted on a holder, connected to a server for traffic generation and data extraction, and powered by a battery. All these pieces of equipment were placed inside a cart and carried around the campus to perform cell coverage measurements.\\
The usage of a private core network allowed us to ease practical issues usually encountered by similar contributions. We used a set of ad-hoc SIM cards that could access the network, removing the need to purchase multiple network subscriptions. Moreover, while most of the related works rely on a speed test website, we installed a speed test tool directly inside the core network. Using remote servers to perform speed tests is a matter of concern because other bottlenecks along the network might alter the results. With our deployment, this obstacle is cleared and the only element impacting our measures is the wireless link.\\
This setup is considered \textit{optimal} since there are no interferents in the area and only one \gls{ue} can connect to the network. In such conditions, the signal strength is directly related to the achievable \gls{mcs} and throughput. Otherwise, the management of more than one user would have originated non-linear relationships between these indicators \cite{where_are_the_cellular_data}. 
Moreover, the choice to deploy a macro base station instead of other architectures makes the outcomes of this campaign very effective for mobile radio network operators since it is the most common network roll-out option and allows co-location with existing facilities.
Milan, as the location for the testbed, is especially fitting to examine urban environments. Parks, trees, and relatively tall buildings characterize the neighborhood and allowed us to study several propagation scenarios, such as the conventional \textit{Line of Sight} and \textit{Non Line of Sight}, urban canyons, and Outdoor-to-Indoor propagation.

\begin{figure*}[!t]
    \begin{subfigure}[b]{.4\linewidth}
        \includegraphics[height = 7.5cm]{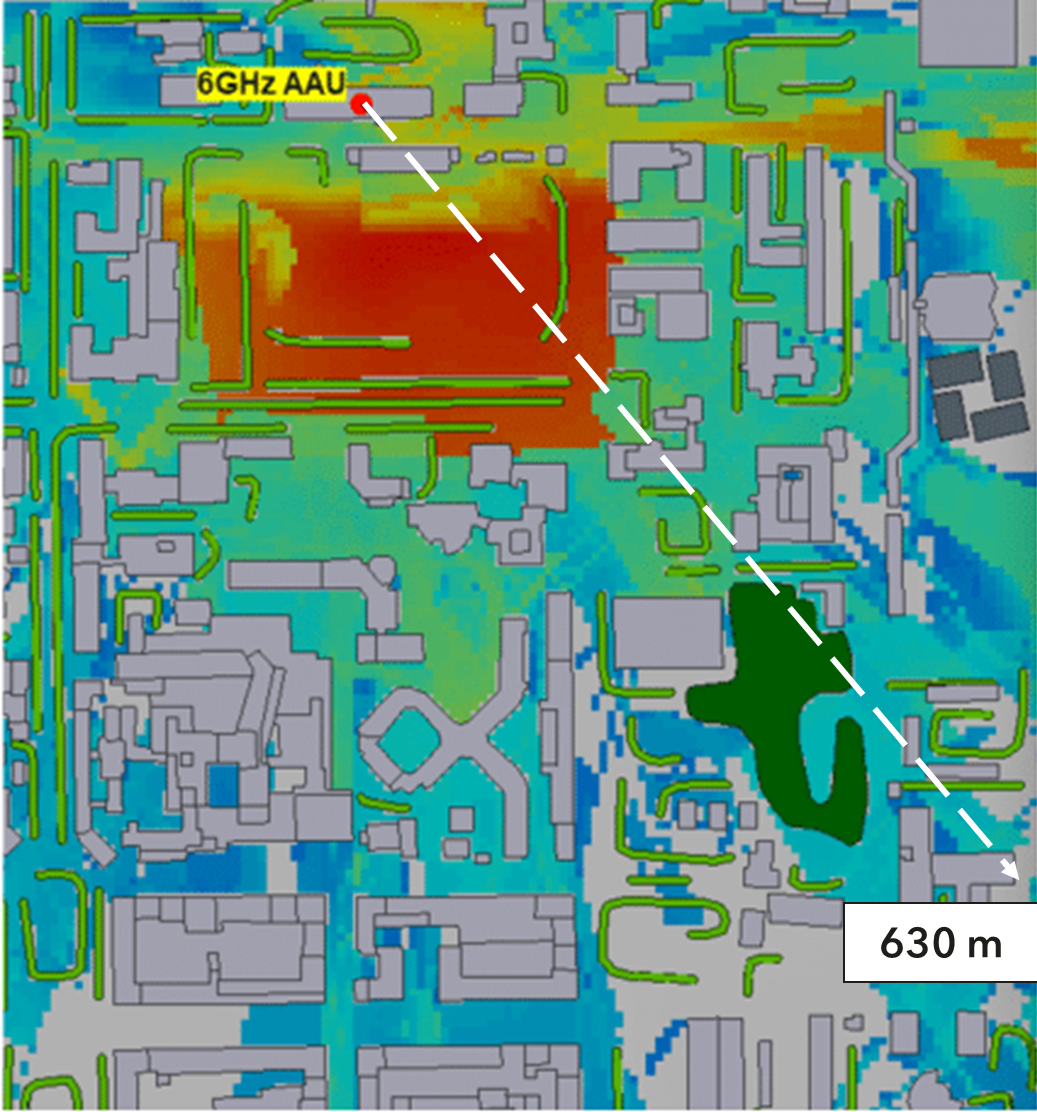}
        \subcaption{\gls{u6g} simulation}
        \label{fig:6GHz_RT}
    \end{subfigure}
    \begin{subfigure}[b]{.4\linewidth}
,        \includegraphics[height = 7.5cm]{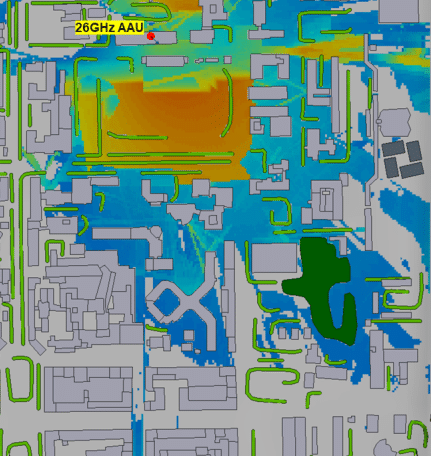}
        \subcaption{mmWave simulation}
        \label{fig:26GHz_RT}
    \end{subfigure}
    \begin{subfigure}[b]{0.1\linewidth}
        \includegraphics[height = 8.5cm]{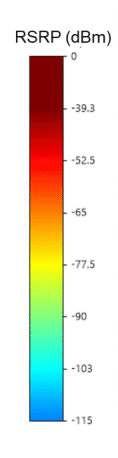}
    \end{subfigure}
    \label{fig:ray_tracing_both}
    \caption{Ray tracing prediction based on 3D vector maps of the Politecnico di Milano premises}
\end{figure*}
\subsection{Ray-tracing simulations}
Before starting the experimental campaign, we acquired the 3D digital maps of the area to perform ray-tracing simulations with the double objective of theoretically verifying the expected coverage as well as driving the selection of interesting positions for the subsequent measurement phase. We used the S\_5GChannel tool by Siradel \cite{siradel} powered by Volcano Urban ray-based model that is able to simulate multiple propagation paths from reflections, diffractions, transmissions and scattering with the objects described by raster and 3D vector layers. In particular, in the software, we placed and oriented the \gls{aau} antenna as in the real testbed and used the realistic radiation pattern of traffic beams to perform ray launching. We carried out point-to-area predictions by placing outdoor measurement points (i.e., user equipment locations) on a bi-dimensional regular grid with a resolution of 5 meters along both \textit{x} and \textit{y-}axis. In such measurement points, we assumed UEs with omnidirectional antenna at 1.5m height.

The results of ray tracing predictions in terms of the \gls{rsrp} heatmaps are reported in \textit{Figure~\ref{fig:6GHz_RT}} and \textit{\ref{fig:26GHz_RT}} for \gls{u6g} and \gls{mmwave}, respectively.
\textit{Figure~\ref{fig:6GHz_RT}} shows \gls{rsrp} peaks of $-55$dBm in the running track placed in perfect \gls{los} with the base station, which gently degrades entering in the urbanized blocks, up to around $600m$ from the base station. At this distance, the \gls{rsrp} decreases to $-115$dBm, considered the minimum power that can be received. Only limited portions of the area are in an outage, mostly placed at the cell edge where buildings shadow the measurement point.\\
\textit{Figure~\ref{fig:26GHz_RT}} follows a behavior similar to the \gls{u6g} simulation but exhibits lower \gls{rsrp} values and a more steeped decrease in \gls{nlos} conditions. The maximum \gls{rsrp} achieved attests to $-60$dBm but sharply drops to less than $-100$dBm when the signal encounters buildings. As a result, the cell radius is smaller. 
However, slightly better coverage is expected in the real implementation of mmWave network thanks to the gain of the \gls{ue} antenna we used.\footnote{We did not consider this effect in our ray-tracing simulations due to the challenges and uncertainties in properly modeling handset antennas.}

\subsection{Upper 6GHz equipment}
The \gls{u6g} network equipment stands out for being a prototype in some of its parts. Specifically, the \gls{aau} and the \gls{tue}, both standard-compliant, were made specifically for a U6G demonstration. On the other hand, the \gls{bbu} is a commercially-available product.
The base station works at a center frequency of 6.8 GHz, on a band of 80 MHz. The \gls{aau} has a gain of 33~dBm and it is equipped with 128 elements for both the transmission and the reception chains.
The \gls{tue} was made of an omnidirectional antenna housed in a commercial smartphone chassis. The antenna was equipped with 4 elements in reception and 2 in transmission, inherently favoring downlink. The radio-frequency front-end was connected to a baseband unit processor, a server for traffic generation, and powered by a generator. All these pieces of equipment were placed inside a cart to reproduce a mobile user.
More details are reported in \textit{Table \ref{tab:hardware_details}}.

\begin{table}
\centering
\caption{Hardware specifications and 5G NR parameters}
\label{tab:hardware_details}
\begin{tabular}{|l||l|l|} 
\hline
\textbf{Parameter}               & \textbf{U6G}                                      & \textbf{mmWave}           \\ 
\hline\hline
AAU coordinates (lat,lon,h)      & \multicolumn{2}{c|}{45.478671, 9.232550, 22 m}              \\ 
\hline
AAU azimuth, down tilt  & \multicolumn{2}{c|}{135°, 2°}                               \\ 
\hline
Center frequency        & 6.8 GHz (n104)                           & 27.2 GHz (n257)  \\ 
\hline
Channel bandwidth~      & 80 MHz                                   & 200 MHz          \\ 
\hline
Subcarrier spacing      & 30 kHz                                   & 120 kHz          \\ 
\hline
Frame structure         & \multicolumn{2}{c|}{TDD 4:1 (DDDSU)}                        \\ 
\hline
Max QAM order (D/U) & \multicolumn{2}{c|}{256/64}                                 \\ 
\hline
AAU TX power            & 37 dBm                                   & 37.5 dBm         \\ 
\hline
AAU gain                & 33 dBi                                   & 32.5 dBi         \\ 
\hline
AAU EIRP                & 70 dBm                                   & 70 dBm           \\ 
\hline
AAU MIMO                & 128T-128R                                & 8T-8R            \\ 
\hline
UE gain                 & $~ 0$ dBi (isotropic) & 20 dBi           \\       
\hline
UE EIRP                 & 22 dBm                                   & 45 dBm           \\ 
\hline
UE MIMO                 & 2T-4R                                    & 2T-2R            \\
\hline
\end{tabular}
\end{table}

\subsection{Millimeter-wave equipment}
The \gls{mmwave} network is fully standard-compliant and commercially available. 
The \gls{ran} consists of one Huawei HAAU5323, interfaced through a 25Gbps eCPRI fiber fronthaul to the baseband unit BBU5900 from the same vendor. The BBU is then connected to the virtualized 5G \gls{cn}. A commercial \gls{cpe} was used as a mobile terminal. More details regarding the mmWave hardware are reported in \textit{Table \ref{tab:hardware_details}}.
The \gls{cpe} was mounted on a holder in the cart, powered by a power bank with AC output, and connected to a laptop. The laptop was extracting the measurements seen by the \gls{cpe} through a drive test log software (Keysight's NEMO) that can access the information available in the \gls{cpe}'s chipset. Differently from the U6G \gls{tue}, \gls{cpe} are directional, giving angular resolution to this campaign.
        

\subsection{Measurement methodology}
The measurement methodology is shared between the two campaigns. 
To carry out the tests, the User Equipment was brought to a measurement point, and two speed tests were launched (both in DL and UL, subsequently). Measurements of the aforementioned \acrlong{kpi} were captured in 15-second windows.
The communication data, ranging from the application to the physical layer, were captured and recorded by the log software. 
The antenna radiation pattern dictates one major difference in the data collection. With the U6G UE antenna, there was no need to test the TUE over different orientations, given its omnidirectional pattern. Oppositely, the mmWave CPE antenna is directive, so a full capture of one point requires orienting the antenna in more directions. In particular, we chose to point them toward the four cardinal points. In the results figure, these directions are indicated by an arrow. The directional pattern increased the angular resolution but required repeating the speed test procedure for each direction.

\section{6GHz} 
\label{sec:6GHz}
In the following, we will present an assessment of the upper 6 GHz 5G NR deployment's performance in both outdoor and indoor environments. Our evaluation encompasses the following metrics: 
\gls{rsrp}, uplink/downlink throughput (measured at the application layer), and statistics related to the active Modulation and Coding Scheme (MCS).

\begin{figure*} 
     \begin{subfigure}[b]{.465\linewidth}
    \includegraphics[width=\linewidth]{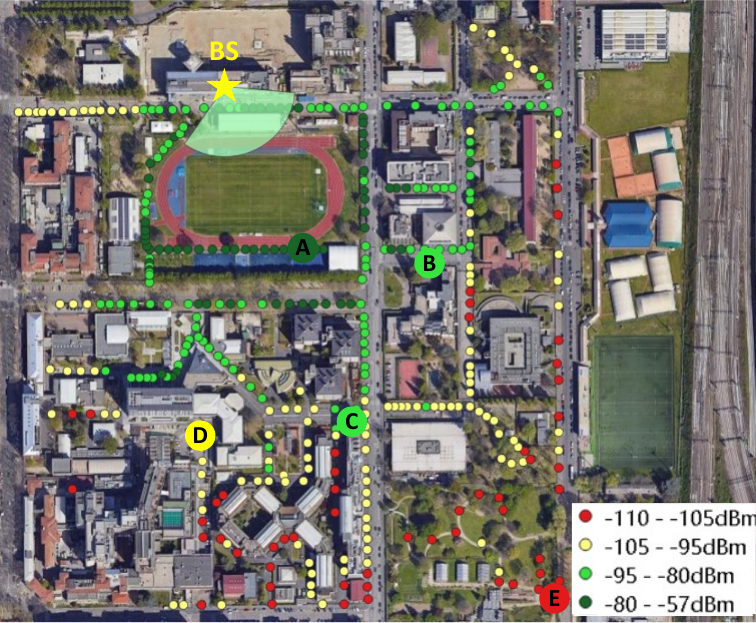}
    \caption{Outdoor downlink RSRP map}
    \label{fig:outdoor_rsrp:rsrp_overlay}
    \end{subfigure}
    \begin{subfigure}[b]{.49\linewidth}
    \includegraphics[width=\linewidth]{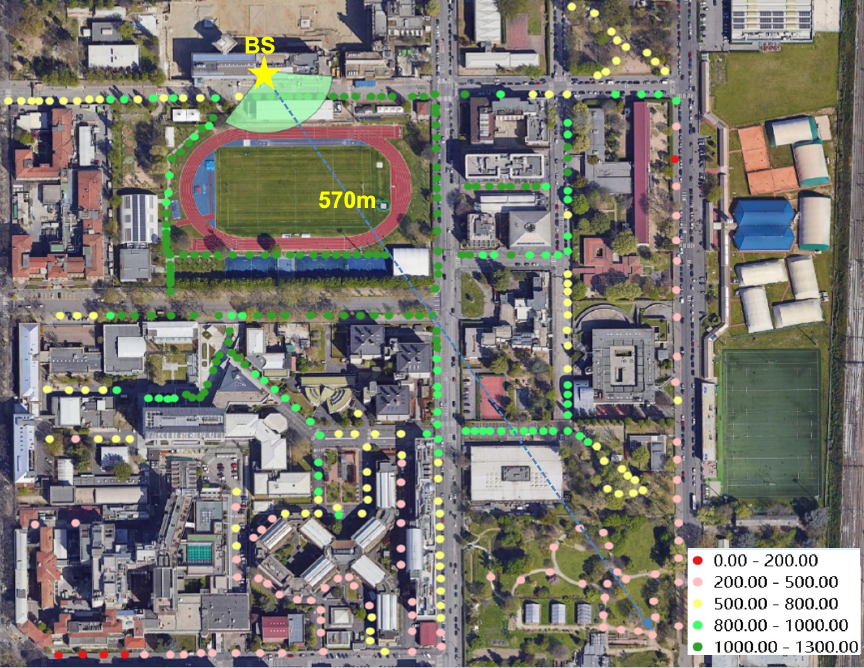}
    \caption{Outdoor downlink throughput map}
    \label{fig:outdoor_rsrp:tp}
    \end{subfigure}
    \caption{Outdoor downlink RSRP and throughput measurements.}
    \label{fig:outdoor_rsrp}
\end{figure*}

\subsection{Outdoor-to-outdoor performance}
\begin{figure*}
    \begin{subfigure}[b]{.3\linewidth}    
    \includegraphics[width=\linewidth]{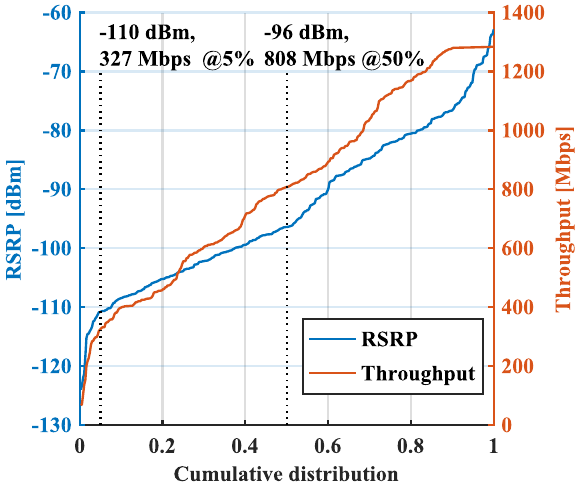}
        \caption{RSRP and downlink throughput}
        \label{fig:rsrp_thr_cdf}
    \end{subfigure} \hfill
    \begin{subfigure}[b]{.3\linewidth}    
        \includegraphics[width=\linewidth]{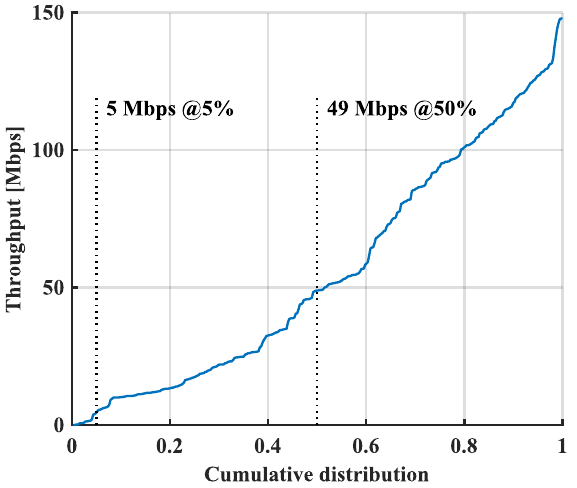}
        \caption{Uplink throughput}
        \label{fig:ul_thr_cdf}
    \end{subfigure} \hfill
    \begin{subfigure}[b]{.3\linewidth}    
        \includegraphics[width=\linewidth]{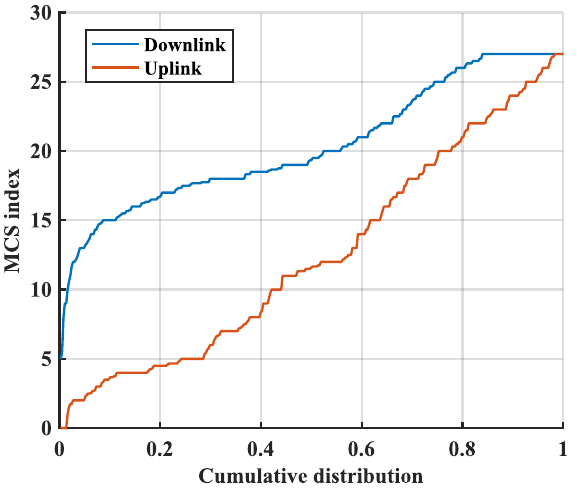}
        \caption{Downlink and uplink MCS}
        \label{fig:mcs_dl_ul}
    \end{subfigure}
    \caption{Empirical cumulative distribution function of network parameters}
\end{figure*}

We start our analysis by showcasing the downlink \gls{rsrp} measurements obtained by the TUE within the outdoor pedestrian zone of the testing area. \textit{Figure \ref{fig:outdoor_rsrp:rsrp_overlay}} visualizes these measurements superimposed on an aerial view of the testing region. Notably, we can observe that test points benefiting from an unobstructed \gls{los} connection with the \gls{bs} experience the highest \gls{rsrp} values, aligning with our expectations.
Interestingly, \gls{nlos} conditions do not necessarily translate to diminished \gls{rsrp} values. Relatively robust \gls{rsrp} measurements were recorded even at test locations where a single building partially obscured the \gls{los}. However, with multiple buildings blocking the \gls{los}, performance decreases rapidly starting from 300~m from the \gls{bs}, as for \textit{points C} and \textit{D}. On the map, below \textit{point C} we can observe a \textit{street-canyon} effect that increases the received signal strength to relatively high values even with multiple blockages and up to the cell edge.\\ 
The \gls{rsrp} values translate in the downlink peak throughput shown in \textit{Fig. \ref{fig:outdoor_rsrp:tp}}. Here it is shown how, even where the signal strength is at its lowest, the peak supported throughput remains well above 200 Mbps. 
In \textit{Fig. \ref{fig:rsrp_thr_cdf}} a quantitative analysis is given through the cumulative distribution functions of measured \gls{rsrp} and downlink throughput. The results illustrate that even with an \gls{rsrp} as low as -110~dBm, a common value found at the cell edge, a peak downlink throughput of approximately 330~Mbps is achievable. It's important to note that these values were only recorded at fewer than 5\% of the test points, while more than 50\% of the test points attain peak throughputs exceeding 800~Mbps.

To discuss uplink performance, consider \textit{Fig. \ref{fig:ul_thr_cdf}}, where the empirical \gls{cdf} for the measured uplink throughput is presented. 
In this context, it is evident that uplink throughput values are generally lower in comparison to the downlink. This discrepancy was anticipated and can be attributed to three main factors. One contribution is the uplink-unfavorable 4:1 Time-Division Duplex (TDD) radio frame configuration. Then, the MIMO capabilities of the Test User Equipment (TUE) are more limited in the uplink, as outlined in \textit{Table~\ref{tab:hardware_details}}. Moreover, the reduced transmit power of the TUE in the uplink further impacts performance. This trend is clearly illustrated in \textit{Fig. \ref{fig:mcs_dl_ul}}, where the empirical \glspl{cdf} of the active \gls{mcs} index in both the downlink and uplink are depicted. As the uplink transmission power diminishes, the RSRP is likewise reduced, resulting in consistently lower supported MCS indices compared to the downlink direction. 

In summary, the system exhibits a degradation in uplink performance, particularly in challenging propagation conditions. This experiment suggests that enhancing uplink performance may require higher transmission power and additional MIMO layers for user equipment. However, this may pose challenges, especially in mobile terminals with power consumption constraints. Alternatively, the use of frame structures that prioritize the uplink can help compensate for the reduction in spectral efficiency, potentially through the implementation of dynamic Time-Division Duplex (TDD)  techniques to prevent unnecessary degradation in downlink performance (e.g., TDD 2:3)~\cite{kim2020dynamic}. 

We conclude this section of the analysis by offering a more detailed examination of the system's performance in a range of distinct scenarios. These scenarios correspond to specific test points, highlighted in \textit{Fig. \ref{fig:outdoor_rsrp:rsrp_overlay}}, which exhibit noteworthy characteristics from a propagation environment perspective.  The test point details and measurements are summarized in \textit{Table~\ref{tab:test_points}}.
\begin{table}[]
\centering
\begin{tabular}{|l|l|l|l|l|l|}
\hline
\multicolumn{1}{|c|}{\textbf{TP}} & \multicolumn{1}{c|}{\textbf{Dist.}} & \multicolumn{1}{c|}{\textbf{LoS}} & \multicolumn{1}{c|}{\textbf{DL TP}} & \multicolumn{1}{c|}{\textbf{UL TP}} & \multicolumn{1}{c|}{\textbf{Rank (D/U)}} \\ \hline\hline
A                                 & $175\,$m                               & yes                               & $1282\,$Mbps                        & $132\,$Mbps                         & 4/2                                      \\ \hline
B                                 & $236\,$m                               & no                                & $992\,$Mbps                         & $95\,$Mbps                          & 4/2                                      \\ \hline
C                                 & $310\,$m                               & no                                & $770\,$Mbps                         & $55\,$Mbps                          & 4/2                                      \\ \hline
D                                 & $344\,$m                               & no                                & $550\,$Mbps                         & $12\,$Mbps                          & 3/1                                      \\ \hline
E                                 & $570\,$m                               & no                                & $332\,$Mbps                         & $6\,$Mbps                           & 2/1                                      \\ \hline
\end{tabular}       
\caption{Specific test point details and measurements.}
\label{tab:test_points}
\end{table}\\

    \begin{figure*}[!ht]
    \centering
    \begin{subfigure}[b]{.4\textwidth}
    \begin{overpic}[scale=0.38, trim= 3cm 0cm 0cm 0] {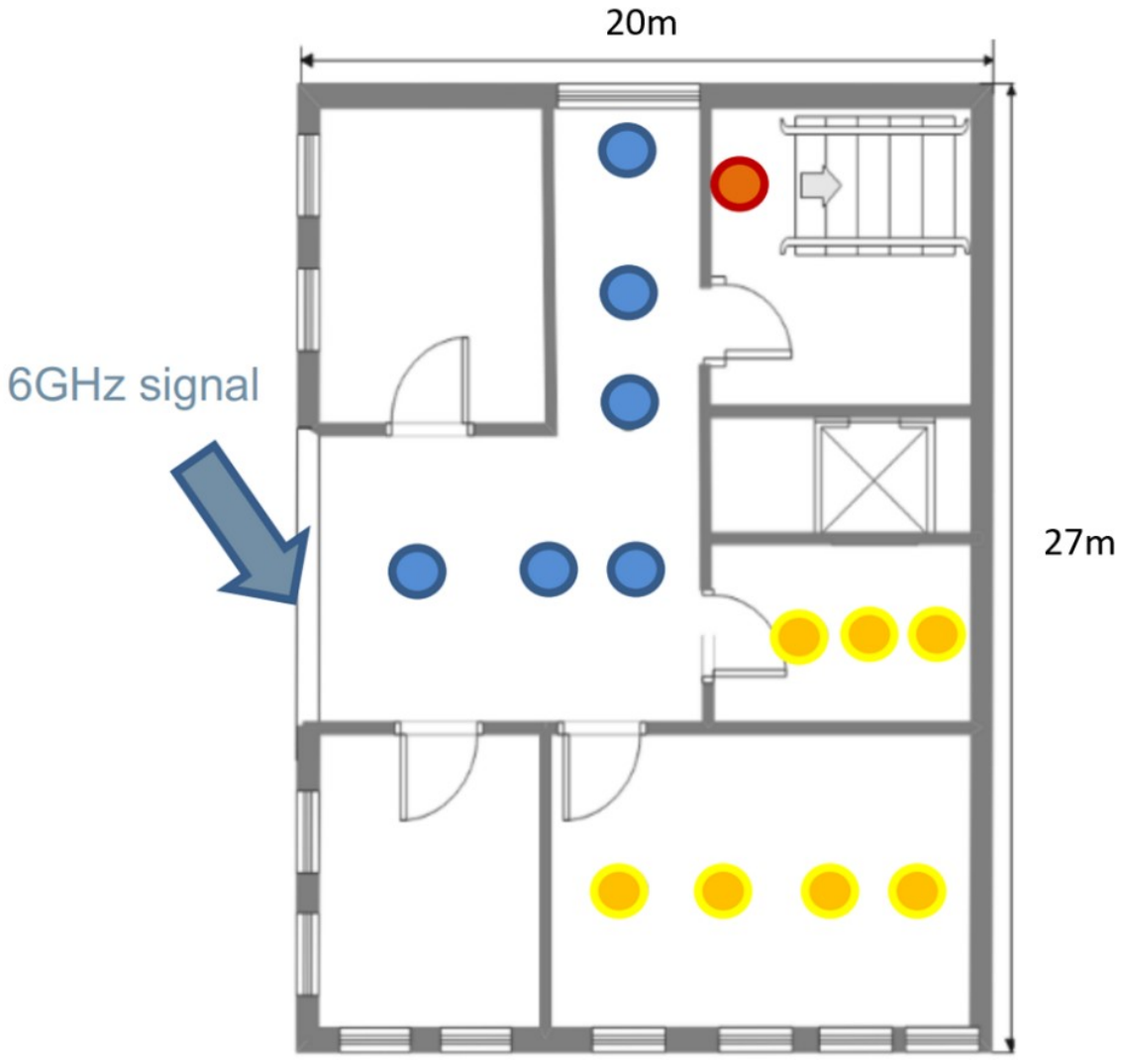}
    \end{overpic}
    \caption{Test points inside the building.}
    \label{o2i:test_points_overlay}
    \end{subfigure}\qquad
    \begin{subfigure}[b]{.5\textwidth}
    \includegraphics[width=\textwidth]{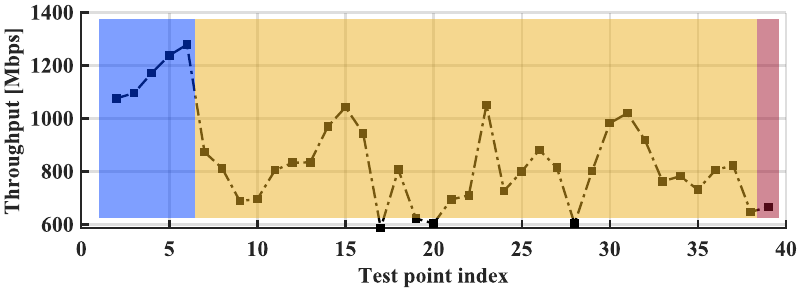}
    \caption{Downlink throughput.}
    \label{o2i:dl_tp}
    \vspace{2ex}

    \includegraphics[width=\textwidth]{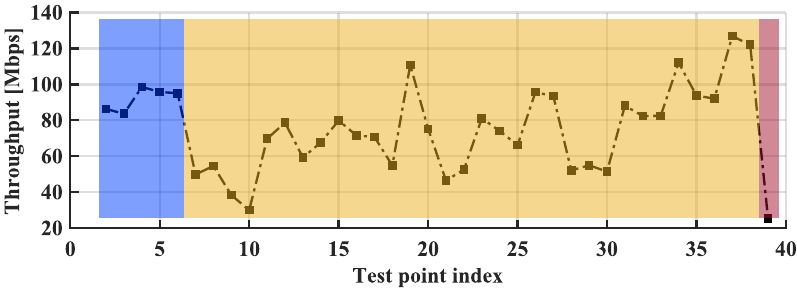}
    \caption{Uplink throughput.}
    \label{o2i:ul_tp}
    \end{subfigure}
    \caption{O2I measurements. Blue, yellow, and red denote test points in the lobby, rooms with windows, and rooms without windows, respectively.}
    \end{figure*}

Test point A is situated 175 meters from the base station and enjoys a clear Line-of-Sight (LoS) connection, making it one of the test points with the most favorable propagation conditions. Here, the downlink throughput reaches 1282 Mbps with a channel rank of 4, while the uplink throughput is 132 Mbps with a channel rank of 2. These throughputs set the performance benchmark for this study. 
Test point B is positioned slightly farther from the base station, at 236 meters. However, it is no longer within the LoS field, as it is obstructed by one building. In this scenario, the downlink and uplink throughput values decrease by about 22\% to 28\% compared to the benchmark, reaching 992 Mbps and 95 Mbps, respectively. The channel rank achieves the maximum in both directions. This drop in performance represents the typical degradation experienced when a single building obstructs the \gls{los}, which is consistent with other test points exhibiting similar propagation characteristics.\\
Test points C and D are situated at slightly more than 300 meters from the base station and both have the \gls{los} blocked by two buildings. However, they experience significantly different performances due to the "urban canyon" effect. Test point C is located on a road flanked by buildings and achieves performances of 770 Mbps for downlink and 55 Mbps for uplink, with no decrease in channel rank. In contrast, test point D is located in a "cul-de-sac" and does not benefit from the same "urban canyon" effect. Here, the throughput measurements are 550 Mbps in downlink and 12 Mbps in uplink, with both channel ranks reduced by 1. This scenario gives clues on the impact of building layout.\\
Lastly, we detail the performance observed at test point E. This spot is placed at the cell's edge, 570 meters away from the base station, with \gls{los} obstructed by multiple buildings. Furthermore, it is situated in a park, devoid of a strong "urban canyon" effect in this case. Ultimately, this represents one of the less favorable propagation scenarios. The measured performance here is 332 Mbps for downlink throughput and 6 Mbps for uplink throughput, with channel ranks reduced to 2 and 1, respectively. While the peak downlink data rate might still accommodate most applications, the uplink data rate appears to be more adversely affected, as previously noted in the general analysis.

\subsection{Outdoor-to-indoor performance}
To evaluate the Indoor-to-Outdoor (O2I) performance, we conducted a similar analysis inside the building highlighted in \textit{Fig.~\ref{fig:aerial_view_with_equipments}}. This building enjoys a \gls{los} connection with the \gls{bs} and is approximately 200 meters far from it. The LoS condition is perfect except for a minor coverage of the facade due to vegetation. Just outside of the building, the downlink and uplink throughputs measure around 1200 Mbps and 130 Mbps. 

\textit{Fig. \ref{o2i:test_points_overlay}} illustrates a selection of crucial test points overlaid on the floor plan of the building's first floor. \\
Test points marked in blue are distributed in the lobby area, where a large glass window offers a direct LoS view with the BS. These points benefit from the best propagation characteristics, as electromagnetic waves only need to penetrate the glass windows. Consequently, both downlink and uplink performances are strong, as depicted in \textit{Figures~\ref{o2i:dl_tp}} and \textit{~\ref{o2i:ul_tp}}, with average values closely resembling those measured outside the building. This confirms the signal's glass penetration capabilities.

Test points marked in yellow are situated in rooms that still have windows, but these do not overlook the BS. They enable radio signals to reach the \gls{tue} with relatively high strength through reflection and other propagation effects.\\
The red test point is located in a room without any windows but with a door opened over the lobby. Here, results demonstrate a significant decrease in performance, with downlink throughput halved and uplink throughput reduced by approximately 80\%. 

As for the previous analysis, it's evident that uplink performance is more sensitive to particularly challenging propagation conditions. Nonetheless, the system demonstrates commendable overall outdoor-to-indoor performance. 
An essential role in the propagation is played by the presence of windows, even when they are placed in NLoS with the base station. 

\section{mmWave}
\label{sec:mmWave}
This section reports the results of the measurement campaign at mmWave. Heatmaps and \gls{cdf} will help provide both qualitative and quantitative samples of millimeter-wave network behavior around the deployment area. Through the beam identifiers, we were also able to reconstruct the approximate path of the beams toward the reception points. This information is also discussed in the most relevant cases.  
A glimpse of outdoor to indoor propagation is also provided at the end of the section.

\subsection{Outdoor-to-outdoor performance}
\begin{figure*}
    \centering
    \includegraphics[width=1.2\columnwidth]{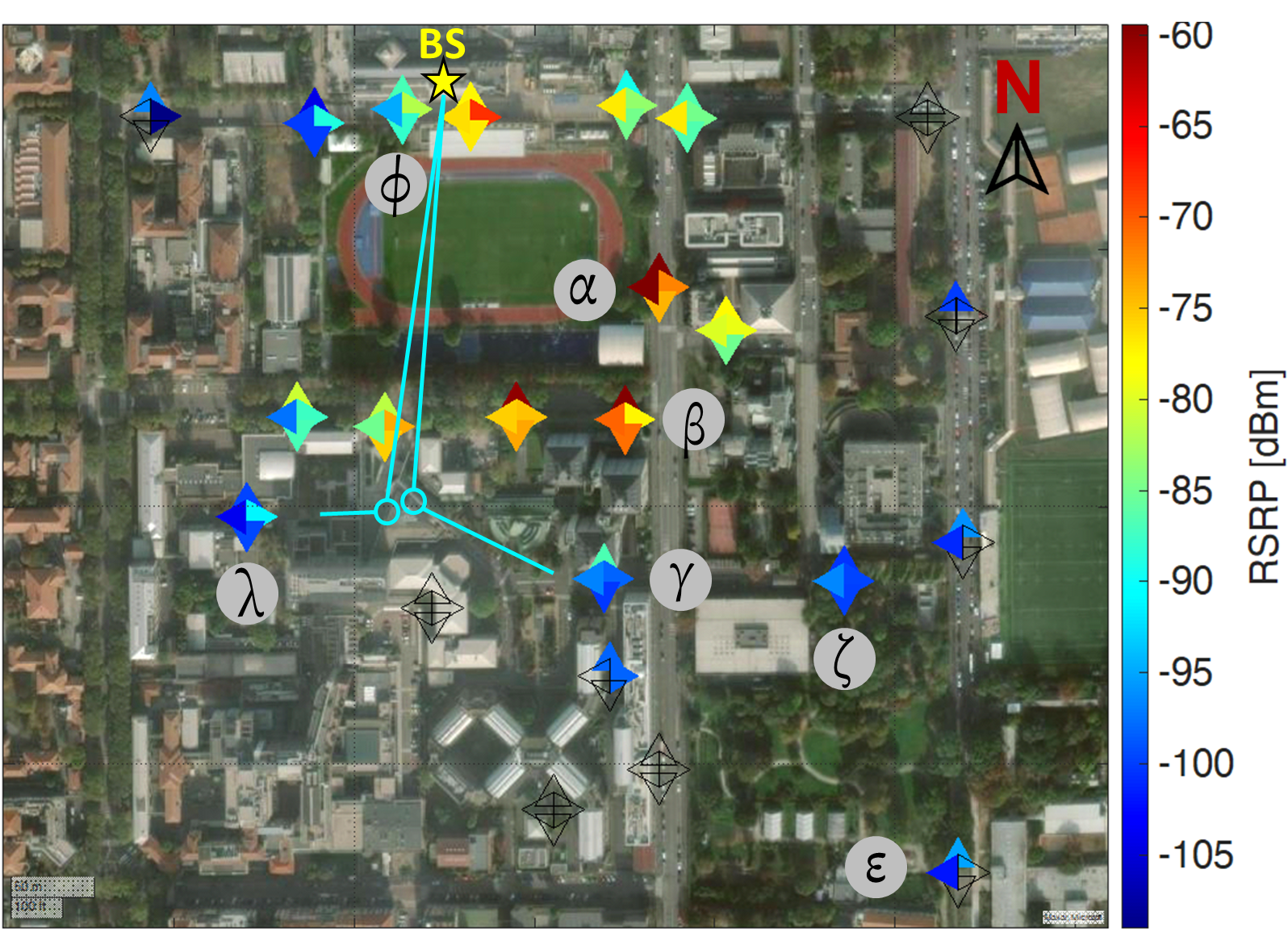}
    \caption{Reference Signal Received Power heatmap}
    \label{fig:heatmap_rsrp}
\end{figure*}

Consider the \gls{rsrp} superimposed to the test area in \textit{Fig. \ref{fig:heatmap_rsrp}}. This parameter is an indicator of the strength of the signal received by the \gls{ue}.

The highest received power occurs under the direct sector illumination. The spots in this region (e.g, $\alpha$, $\beta$) reach \gls{rsrp} values in the order of -60 dBm, which sets as the maximum received signal strength and corresponds to peak throughputs of slightly less than 1.3Gbps in \gls{dl} and 250 Mbps in \gls{ul}. Once again, this unbalance in favor of \gls{dl} throughputs originates from the \gls{tdd} 4:1 slot configuration, from the lower transmission power of the \gls{cpe} with respect to the \gls{bs} and the maximum QAM order. The pointing direction is also relevant. Point $\alpha$ in the north and west direction stably connects to a beam enjoying $-59$ dBm and decreases to around $-73$ dBm when pointing east and south.
Point $\beta$ overlooks a street skirted by trees that were mostly bare at the time of the campaign. Here a direct \gls{los} connection is established with the \gls{bs} on the rooftop. In the most favorable direction, the RSRP is equal to the above-mentioned one, with a lower peak towards the west, probably due to the absence of close buildings that can provide major reflections.  

The points immediately under the base station, shadowed by the building hosting the site, do not enjoy a direct \gls{los} link. Nevertheless, a strong urban canyon effect greatly improves the coverage.  
For example, spot $\phi$, confined between two buildings, reaches performances that coincide with the perfect LoS of point $\alpha$, thanks to the aforementioned effect. However, moving a few meters west, where buildings do not surround the road, is sufficient to make this effect fade, and the signal rapidly worsens. The beam choice (not shown in the figure) in the point exactly under the \gls{bs} is distinctive: the best beam quickly switches every time the CPE's orientation is changed. This is due to the separation between beams being not pronounced as a consequence of the reflection, thus creating a crowded scene where beams have similar propagation conditions. 
On the top-right side of the image, the signal rapidly decreases, denoting the sector's edge.

On the center-left part of the image, above $\lambda$, we enter a \textit{soft} \gls{nlos} condition, where signals as high as -74 dBm are received from reflections, as the south and west-pointing arrows suggest. \\
Shifting toward the center of the picture, a dense block of buildings is encountered. 
Interestingly, the signal can infiltrate inside the block and reach locations where the \gls{los} is completely blocked. For example, spot $\lambda$ reaches -91 dBm while pointing east, thanks to a strong reflection on the building on its right, also highlighted in \textit{Figure \ref{fig:heatmap_rsrp}}.\\  
On the opposite side of the block, the spot $\gamma$ gets its maximum power when pointing north. Given the inability of the beam to pass through the building ahead, the best beam must bounce on the building on the left and is eventually reflected, as highlighted in the figure.\\
Point $\zeta$ is also placed in soft \gls{nlos}: the direct path is covered by a 12m-tall building and foliage. Still, the connection is kept in every direction, reached by a decent signal strength which translates into downlink throughputs of 407 Mbps. 

The presence of the spots on the right side of the picture is surprising. The existence itself of a connection in these points is notable since the ray tracer simulation foresees RSRP values as low as $-115$ dBm or even no coverage. Point $\epsilon$ is placed at slightly less than 600m, at the edge of the 6GHz cell, and its direct path is covered by several buildings and trees. The signal here is very low, indeed it cannot be received in every direction. Still, one orientation reaches an RSRP of  $-100$dBm, which can supply a peak throughput of 435Mbps in DL. Part of the merit belongs to the antenna of the \gls{cpe}, which gain helped in extracting the \gls{mmwave} signal out. The uplink transmission in this area is afflicted by the harsh propagation, demonstrating once again that the UL is more delicate.
More details regarding some of these peculiar test points are reported in \textit{Tab. \ref{tab:test_points_mmwave}}. \\
Finally, some spots where the signal cannot reach the UE are displayed with a black empty arrow. 

\begin{figure}[t!]
    \centering
    \includegraphics[width=\linewidth]{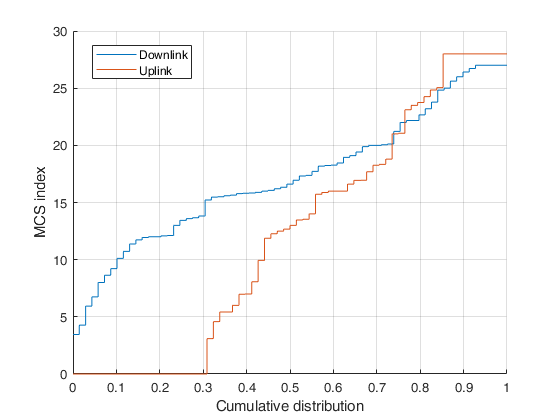}
    \caption{Downlink and uplink Modulation and Coding Scheme \gls{cdf}}
    \label{fig:mcs_cdf}
\end{figure} 
Fig. \ref{fig:mcs_cdf} reports the \gls{mcs} cumulative distribution function for both downlink and uplink. 
Note that these MCS indexes originate from two different MCS tables for uplink and downlink, due to the hardware configuration. 
The amount of locations where only downlink transmission is withstanded impacts the distribution. Around 30\% of the points where the CPE can attach to the base station can transmit in the downlink but do not support the uplink transmission (represented with MCS equal to 0). 
Interestingly, around 15\% of the UL measurements could reach the highest MCS, which is around twice the amount that can achieve the same in DL. This behavior is dictated by the specific implementation of the hardware used.  
The number of missed connections makes the UL \gls{mcs} curve steeper.
Interpreting this result, it could be argued that the UL connection is more fragile since the margin between maximum performance and a missed connection is small.
\begin{figure}[]
    \centering
    \includegraphics[width=1\linewidth]{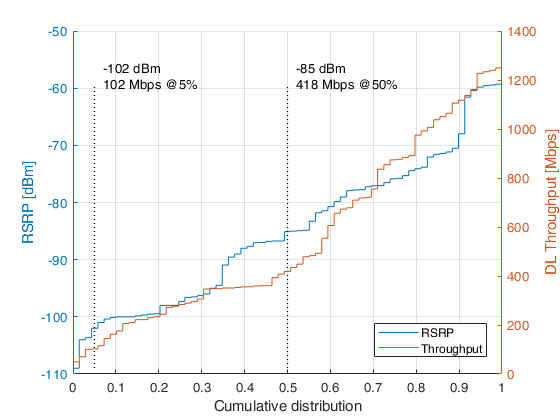}
    \caption{Downlink and RSRP \gls{cdf}}
    \label{fig:dl_plus_rsrp_cdf}
\end{figure}
\begin{table}[]
\centering
\begin{tabular}{|l|l|l|l|l|l|}
\hline
\multicolumn{1}{|c|}{\textbf{TP}} & \multicolumn{1}{c|}{\textbf{Dist.}} & \multicolumn{1}{c|}{\textbf{LoS}} & \multicolumn{1}{c|}{\textbf{DL TP}} & \multicolumn{1}{c|}{\textbf{UL TP}} & \multicolumn{1}{c|}{\textbf{RSRP}} \\ \hline\hline
$\alpha$                          & $190\,$m                               & yes                               & $1239\,$Mbps                        & $235\,$Mbps                         & $-59\,$dBm                                    \\ \hline 
$\lambda$                            & $285\,$m                               & no                                & $708\,$Mbps                         & $0\,$Mbps                           & $-91\,$dBm                                    \\ \hline
$\gamma$                          & $315\,$m                               & no                                & $355\,$Mbps                         & $17\,$Mbps                           & $-87\,$dBm                                    \\ \hline
$\zeta$                           & $395\,$m                               & no                                & $407\,$Mbps                         & $0\,$Mbps                           & $-96\,$dBm                                    \\ \hline
\large$\epsilon$                            & $568\,$ m                              & no                                & $435\,$Mbps                         & $10\,$Mbps                          & $-95\,$dBm                                    \\ \hline
\end{tabular}       
\caption{Specific test point details and measurements.}
\label{tab:test_points_mmwave}
\end{table}
In \textit{Fig. \ref{fig:dl_plus_rsrp_cdf}} the comparison between downlink throughput and RSRP is reported. The trend is as expected: high RSRP corresponds to high throughput values. This remarks that with our setup, the speed of the connection can be in general inferred by the reference signal strength.

From the results, some general trends can be derived.  
The angular resolution of the setup gives clues on the relevance of reflections to get the signal in NLoS conditions. Points reported in \textit{Fig. \ref{fig:heatmap_rsrp}} show that outside the LoS-illuminated sector, the highest-quality signal often arrives from reflections.
One main difference between the LoS and NLoS conditions is the variance of the throughput values during the 15s-long capture. While the signal power is rather stable in LoS, the throughput and RSRP values fluctuate significantly in NLoS, as well as the anchor beam. 
\vspace{-0.1cm}
\subsection{Outdoor-to-indoor performance}
\begin{figure}          
    \centering
    \includegraphics[width=1\linewidth]{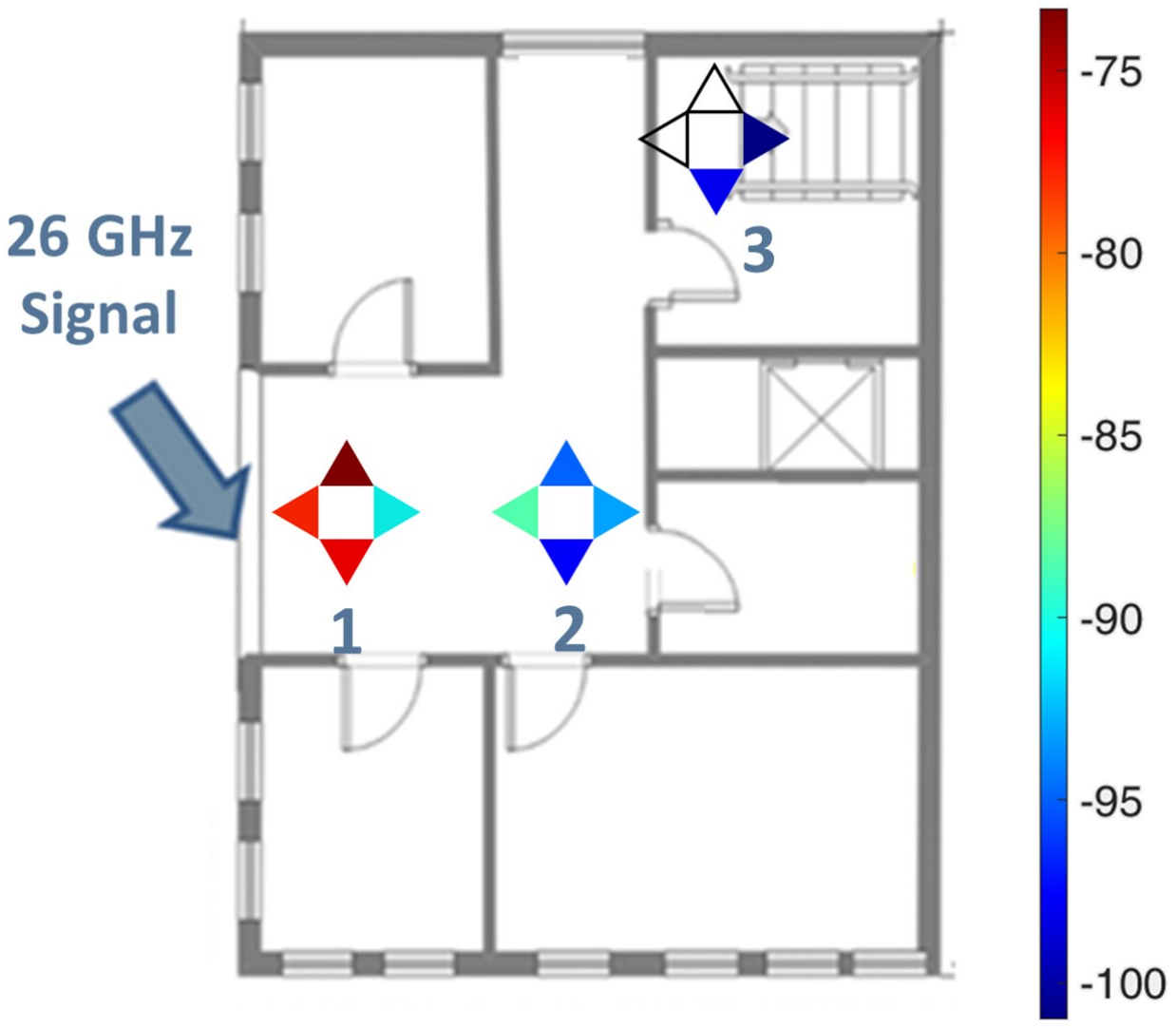}
    \caption{Outdoor to indoor RSRP results [dBm]}    
    \label{fig:26GHz_o2i}
\end{figure}

The campaign is completed with indoor measurements, shown in \textit{Fig. \ref{fig:26GHz_o2i}}. Those measurements were captured in the same building discussed in the previous section (highlighted in Fig. \ref{fig:aerial_view_with_equipments}). The entrance is made of a two-layer glass window. 
\textit{Point 1} was chosen close to the door, to observe its effect on the signal. The peak throughput does not vary from what was obtained outside of the same building, indeed it reaches the upper limit peak of 1.3Gbps in DL and 251Mbps in UL towards the optimal direction (west). However, this speed fluctuates more than the corresponding measurement taken just outside of the glass, and the RSRP tops at $-73$dBm, which means a 14 dBm loss.
This gives clues regarding the absorption of the window. Comparing \textit{point 1} north and east directions, one can also conclude that the signal reflected inside the lobby is 17 dBm lower than the direct path.\\
\textit{Point 2} has less favorable conditions thus, as expected, the throughputs are slightly lower, reaching up to 560 Mbps in DL and 70 in UL. \\ 
\textit{Point 3} is instead placed in a room with no windows but with a door open on the lobby. Here the CPE only manages to connect in two directions (those closest to the door), where it maintains a good connection in DL with 425 Mbps and keeps the link with 20 Mbps in UL. 
These measurements once again highlight how the presence of windows positively impacts indoor penetration, while the concrete completely blocks the signal. It is worth noting that in almost every connectable direction, the downlink throughput exceeds 400Mbps throughput. 



\vspace{-0.1cm}
\section{Comparison}
\label{sec:comparison}



 
The measurement campaigns summarized in the previous sections prove how both U6G and mmWave have the potential of bringing Gigabit-level performance to the \gls{ran}. However, the two deployments still retain some fundamental differences. Exploiting the co-location of the two cells, in this section we give a comparative analysis of both systems' performance to provide a comprehensive understanding of their respective strengths and limitations. 

\vspace{-0.2cm}
\subsection{Expected vs measured throughput and overhead}
The U6G and the mmWave cells present the maximum achievable throughput values reported in Tab.~\ref{tab:max_throughput_formula}. The value for \gls{u6g} was computed using the \gls{3gpp} formula in \cite{throughput_formula}, while, for \gls{mmwave}, the \gls{3gpp} formula in \cite{throughput_formula_TBS} was used. This choice was obliged since \cite{throughput_formula} was evidently not precise for high frequencies, and resulted in an unjustifiedly high value.   
These theoretical values are computed considering the cell configuration (i.e. bandwidth, MIMO layers, etc.) and represent the achievable MAC throughput for both deployments. Note that the UE have different capabilities. In the \gls{ul} direction, both cells have the same number of layers and the expected throughput difference is mainly given by the larger bandwidth available at mmWave. On the other hand, the \gls{u6g} cell has 4 MIMO layers in the \gls{dl} direction, two more than the \gls{mmwave} cell, which balances the larger bandwidth gains in mmWave. \\
Tab.~\ref{tab:max_throughput_formula} also reports the maximum measured throughput in our campaigns, showing how these values are slightly lower than the theoretical maximum. This result was expected, and it is known to be caused by several factors. 
First, it is not easy to estimate the overhead caused by the control plane. \gls{3gpp} suggests empirical values that account for it, but some margin remains. In realistic settings, the control plane overhead is tightly related to both the cell configuration (e.g. PRACH, channel estimation and positioning signals density) and the instantaneous network conditions (e.g. handovers, user attachments). Thus, a mismatch is to be expected.
Second, this formula computes the MAC throughput but does not consider additional bottlenecks generated by higher layers. Among these, the transport layer is most notable, as it is known that TCP under-performs over less-than-stable wireless links, especially \gls{mmwave}~\cite{REN202180}.
Overall, the gap between theoretical and measured results can be attributed to reasonable deteriorations related to the system implementation and the real-world evironment.  

We observed this behavior for both \gls{u6g} and \gls{mmwave}, but the effects are slightly more pronounced for the latter. This is, once again, expected due to the harsher propagation, the increased control plane overhead, and the lower TCP-over-\gls{mmwave} performance. 

\subsection{Performance and coverage comparison}
\begin{table}
    \centering
\begin{tabular}{l|c|c|c|c|} 
\cline{2-5}
                   & \multicolumn{2}{c|}{\textbf{U6G}}                     & \multicolumn{2}{c|}{\textbf{mmWave}}                  \\ 
\hhline{~====|}
                   & UL                        & DL                        & UL                       & DL                         \\ 
\cline{2-5}
Theoretical [Mbps] & 170.33                    & 1379.94                   & 253.74                   & 1354.34                       \\ 
\cline{2-5}
Measured [Mbps]    & \multicolumn{1}{l|}{132}  & \multicolumn{1}{l|}{1282} & \multicolumn{1}{l|}{235} & \multicolumn{1}{l|}{1239}  \\ 
\cline{2-5}
Loss [\%]          & \multicolumn{1}{l|}{22.5} & \multicolumn{1}{l|}{7.1}  & \multicolumn{1}{l|}{7.4}  & \multicolumn{1}{l|}{8.5}  \\
\cline{2-5}
\end{tabular}
    \caption[Maximum achievable throughput in Mbps computed with 3GPP formula \cite{throughput_formula}]{Maximum achievable throughput in Mbps computed with 3GPP formula~\cite{throughput_formula} and maximum measured values.}
    \label{tab:max_throughput_formula}
\end{table}
We continue the analysis by comparing the performance and coverage of both systems at different test points, which we report in Tab. \ref{tab:comparison}. The two technologies show fundamental differences in the cell configurations and device capabilities, such as bandwidth, numerology and \gls{mimo} layers. While these differences are representative of typical cell deployments, they strongly impact the final performance. As such, alongside a direct throughput comparison, we also report the spectral efficiency computed in terms of capacity over Hertz per \gls{mimo} channel. Furthermore, the values reported for the \gls{mmwave} case are taken selecting the best pointing direction for the \gls{mmwave} \gls{ue}.

\begin{table*}  
    \centering
\begin{tabular}{|c|c|c|cc|cc|cc|cc|}
\hline
\multirow{2}{*}{Points} & \multirow{2}{*}{Distance {[}m{]}} & \multirow{2}{*}{LoS} & \multicolumn{2}{c|}{Bandwidth {[}MHz{]}}                        & \multicolumn{2}{c|}{Rank (D/U)}   & \multicolumn{2}{c|}{Throughput (D/U) {[}Mbps{]}} & \multicolumn{2}{c|}{Spectral efficiency (D/U) {[}bps/Hz/ch{]}}                               \\ \cline{4-11} 
                        &                                   &                      & \multicolumn{1}{c|}{U6G}                 & mmWave               & \multicolumn{1}{c|}{U6G} & mmWave & \multicolumn{1}{c|}{U6G}          & mmWave       & \multicolumn{1}{c|}{U6G} & mmWave  \\ \hline
A, $\alpha$             & 175                               & y                    & \multicolumn{1}{c|}{\multirow{5}{*}{80}} & \multirow{5}{*}{200} & \multicolumn{1}{c|}{4/2} & 2/2    & \multicolumn{1}{c|}{1282/132}     & 1239/235     & \multicolumn{1}{c|}{5.4/3.3}                                                       & 4.2/2.3 \\
B, $\beta$              & 236                               & n                    & \multicolumn{1}{c|}{}                    &                      & \multicolumn{1}{c|}{4/2} & 2/2    & \multicolumn{1}{c|}{992/95}       & 891/112      & \multicolumn{1}{c|}{4.2/2.5}                                                       & 3.0/1.2 \\
C, $\gamma$             & 310                               & n                    & \multicolumn{1}{c|}{}                    &                      & \multicolumn{1}{c|}{4/2} & 2/2    & \multicolumn{1}{c|}{770/55}       & 355/17       & \multicolumn{1}{c|}{3.2/1.3}                                                       & 1.2/0.4 \\
D, n/a                  & 344                               & n                    & \multicolumn{1}{c|}{}                    &                      & \multicolumn{1}{c|}{3/1} & n/a    & \multicolumn{1}{c|}{550/12}       & 0/0          & \multicolumn{1}{c|}{3.1/0.4}                                                       & n/a     \\
E, $\epsilon$           & 570                               & n                    & \multicolumn{1}{c|}{}                    &                      & \multicolumn{1}{c|}{2/1} & 2/2    & \multicolumn{1}{c|}{332/6}        & 435/10       & \multicolumn{1}{c|}{2.8/0.4}                                                       & 1.5/0.4 \\ \hline
\end{tabular}
    \caption{Comparison between U6G and mmWave capabilities}
    \label{tab:comparison}
\end{table*}

Points A and $\alpha$ represent the position with the highest measured performance in both campaigns, where both technologies can establish a full-rank \gls{los} connection. Here the lower downlink spectral efficiency and reduced \gls{mimo} layers of \gls{mmwave} are compensated by the larger available bandwidth, making the downlink throughput comparable. On the other hand, the uplink throughput is almost doubled in \gls{mmwave} since 2 \gls{mimo} layers are active in both cells. 

Points B, $\beta$ are in a \gls{nlos} condition caused by a single building. Despite the obstruction, we still observe a relatively high throughput, with a more pronounced impact on the \gls{mmwave} system, as expected. Nonetheless, the \gls{mmwave} uplink is still higher than the \gls{u6g} one. Points C and $\gamma$ are obstructed by multiple buildings, but they benefit from a urban canyon effect, as previously mentioned. Here we observe a sharp \gls{mmwave} performance degradation. 

More in detail, the \gls{mmwave} connection is still at full rank, but the equivalent spectral efficiency is more than halved with respect to the previous case. On the contrary, the \gls{u6g} system experiences a less pronounced performance degradation. Point D represents a particularly harsh test, as multiple tall buildings obstruct the \gls{los} with no urban canyon effect. Here the \gls{mmwave} \gls{ue} fails to attach to the base station, while the \gls{u6g} system can still provide almost half of the full capacity. 

Points E and $\epsilon$ are the furthest from the base station. However, the \gls{los} obstruction here is less severe with respect to the previous case. In this case, with a maximum achievable channel rank of 2 for both system, the higher \gls{mmwave} bandwidth compensates for the reduced spectral efficiency and allows higher performance in both directions. 

According to the detailed comparison given above, some general trends can be observed. Despite the different cell configuration and device capabilities, the two systems show comparable best case performance. As expected, \gls{mmwave} shows higher sensitivity to penetration losses and achieves lower performance than \gls{u6g} under severe \gls{nlos}. On the other hand, the \gls{u6g} system heavily relies on multiple active \gls{mimo} streams to offer a performance level on par with \gls{mmwave}. Consequently, \gls{mimo}-adverse propagation environments can be better exploited by \gls{mmwave}. Indeed, channel separation performs generally better at higher frequencies~\cite{mizmizi2021}, allowing the \gls{mmwave} system to potentially enjoy a higher channel rank. At the same time, a single \gls{mmwave} spatial stream has higher potential throughput with respect to \gls{u6g}, making the loss of spatial diversity less impactful on the \gls{mmwave} system. Such behaviour is expected and it is confirmed by our results in points E and $\epsilon$.

For what concerns the \gls{o2i} scenario, both systems show a good penetration of the building's glass window. As the test point is moved deeper inside the building, the performance degradation follows the same trend for both. Most notably, however, the direction of the \gls{mmwave} \gls{ue} has an impact on the indoor performance, while this is not true for the ominidirectional \gls{u6g} \gls{ue}.

\subsection{Performance improvement margins}
From the comparative analysis given in this section, it appears that \gls{u6g} dominates \gls{mmwave} almost entirely in the context of macro urban coverage. Such result is not surprising, mostly due to the well-known harsher propagation at \gls{mmwave}. However, the cell configuration and the technology maturity level of the involved devices have to be taken into account when forecasting for realistic future performance.

Indeed, in those propagation environments not heavily dominated by penetration losses, \gls{u6g} performs better because up to 4 \gls{mimo} layers can be activated, as opposed to \gls{mmwave} which has only 2. However, statistical data from urban macro coverage at lower frequencies, thus representative of an high technological maturity and realistic \gls{ue} capabilities, shows that up to 2 \gls{mimo} layers are active for most of the connections, even when 4 layers are available~\cite{rochman2023comprehensive}. At the same time, we can expect 4 \gls{mimo} layers being available also for \gls{mmwave} \glspl{ue} with higher technological maturity. Additionally, up to $400$ MHz of cumulative bandwidth are available for \gls{mmwave}, potentially doubling the overall performance. On the other hand, the $80$ MHz bandwidth configuration of our \gls{u6g} deployment is already reasonably close to the maximum bandwidth availability in \gls{u6g}. All together, the suggestion goes towards a larger performance improvement potential for \gls{mmwave} deployments, especially if higher \gls{mimo} capabilities will be made available by improved technological maturity. 

Finally, uplink traffic is gaining increasing attention~\cite{ngmn_uplink_wp} but both systems show overall poor performance. This can be mitigated by selecting a more favourable \gls{tdd} frame structure in those cells where enhanced uplink is required. However, this option is only viable for \gls{mmwave} networks, as \gls{u6g} coverage is expected to operate at the macro level, making heterogeneous frame structures impossible due to inter-cell synchronization constraints.

\section{Related works}
\label{sec:related}
Given the recent introduction of the U6G in the standard and the WRC-23 licensing decision, there are no other articles that document a demonstration of a cellular network working at these frequencies. Therefore, this is the first work that reports measurement data on a U6G 5G deployment.

The same does not hold for mmWave, for which few contributions are instead present in the literature.
Still, the relatively small number of mmWave commercial 5G deployments delayed the measurement effort. At the time of writing, the literature containing such measurements is composed of only a few contributions, briefly commented in the following. This shortage hinders the study and the optimization of such networks, which is one of the impairments for the adoption of this technology.\\
Authors in \cite{a_comparative_measurement} carry out several measurements over commercial, \gls{nsa}, 5G network deployed in Chicago, Illinois, working at 28 and 39 GHz. While a user is moving, the signal's physical data and throughputs are extracted, and emphasis is put on the derived beam management techniques. Their dataset is publicly available\footnote{ \url{https://5gbeams.umn.edu}}.
The work in \cite{a_first_look} studies the first mmWave networks deployed in the U.S., namely in Minneapolis, Chicago and Atlanta in 2019. It offers insights on both stationary and moving UEs, including an analysis of handoffs. Collected physical-layer parameters are limited to the \gls{rsrp}, while some more information is extracted from upper layers. The dataset is publicly available\footnote{\url{https://fivegophers.umn.edu/www20/}}.
Authors in \cite{an_in_depth_study_of_ul} perform measurements in Boston, Chicago, and Indianapolis at 28 and 39 GHz, both static and in mobility, but limited to uplink and with a main focus on upper layers\footnote{\url{https://github.com/NUWiNS/sigcomm-5gmemu-5g-mmWave-uplink-data}}. 
Authors in \cite{Vivisecting_Mobility_Management_in_5G} conducted a large test drive to map the handoffs behavior, passing also through mmWave stations.
Finally, \cite{Cellular_Deployments_in_Chicago_and_Miami} extends the already rich set of campaigns in Chicago with one in Miami, Florida.\\  
This literature produced a significant dataset for few major cities in the U.S. 
The only campaign that took place in Europe is discussed in \cite{sectors_beam_telenor}. Coverage measurements are done in Oslo, Norway, and the results include interesting insights on foliage, human body attenuation, and propagation close to water.\\ 
The lack of measurement campaigns in Europe is an important issue: while it is, in general, difficult to extend propagation characteristics through different environments, this is especially true for European cities, which significantly differ from those of the available campaigns. Therefore, more data and tests are still needed. \\
Finally, a few more contributions focus on more specific aspects of the network operations.
The work in \cite{Outdoor_to_Indoor} reports signal strength, throughput and latency of an outdoor-to-indoor measurement performed in Chicago, also comparing mmWave performances (28 GHz) with LTE ones.
Authors  in \cite{a_variegated_look_at_5G_in_the_wild} measure high-level characteristics and \gls{ue} power consumption in 28/39 GHz 5G networks in the U.S. and compare them to those of 4G in the same location. Their dataset is publicly available\footnote {\url{https://github.com/SIGCOMM21-5G/artifact}}.
Authors in \cite{lumos5g} collected data\footnote{ \url{https://lumos5g.umn.edu/}} and developed a machine learning model to predict throughputs based on them.
Finally, \cite{5G_mmWave_PHY_Latency} focuses on the measurement of latency and other end-to-end performance indicators.

\section{Conclusion}
\label{sec:conclusion}
This paper presents a thorough analysis of the results of two measurement campaigns on 5G networks working in upper 6GHz and millimeter-wave bands. \\
The U6G network, operating at a central frequency of $6.8$ GHz, covered an area with a 600m radius and achieved throughputs of up to 1.3 Gbps in downlink and 150 Mbps in uplink. The network demonstrated stable performance, maintaining throughputs higher than 200 Mbps in downlink for the majority of points, with exceptions due to signal strength issues constituting less than 5\%. The ray-tracer results overall comply with the empirical results. Indoor penetration is larger than expected, particularly in the presence of windows.

The network working at 26GHz exhibits a coverage area compliant with the one predicted by the ray-tracing simulations. Maximum performances achieved are about 1.25 Gbps in downlink and 230 Mbps in uplink, demonstrating the high potential of the mmWave to offer unprecedented uplink speeds in LoS. This technology can also boast a high margin for improvement given the large bandwidth available at those frequencies. Despite the harsher propagation at mmWave, the base station deployment covered an area comparable to the existing commercial macro cells in the neighborhood, working at sub-6GHz frequencies. Outdoor to indoor propagation is limited to spaces where windows are present. 

5G and beyond networks heavily rely on the availability of additional spectrum, and U6G and mmWave bands are relevant assets. This paper showcases the capabilities of radio access networks operating at such frequencies and compares them within a realistic scenario for which only limited information is currently available. 

\subsection{Acknowledgments}
The research in this paper has been carried out in the framework of Huawei-Politecnico di Milano Joint Research Lab. The Authors acknowledge Huawei Milan research center for the collaboration.

The work of E. Moro and I. Filippini was partially supported by the European Union under the Italian National Recovery and Resilience Plan (NRRP) of NextGenerationEU, partnership on “Telecommunications of the Future” (PE00000001 - program “RESTART”, Structural Project 6GWINET).

\bibliographystyle{ieeetr}
\bibliography{content/bibliography}
\begin{IEEEbiography}[{\includegraphics[width=1in,height=1.25in,clip,keepaspectratio]{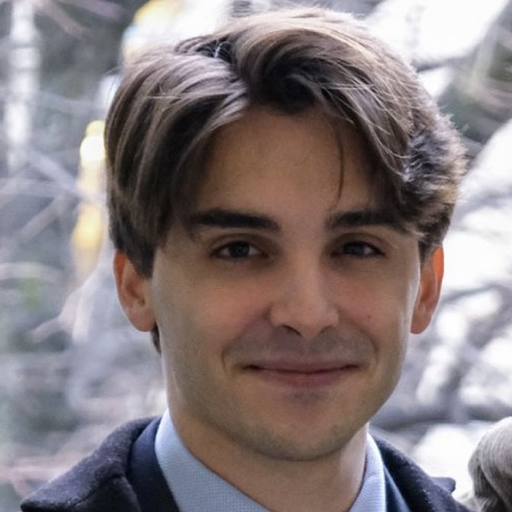}}]{Marcello Morini} (Member, IEEE)  is currently a PhD student at Politecnico di Milano, Department of Electronics, Information and Bioengineering. His research area is Telecommunication, with a focus on high-frequency radio access networks and smart propagation environments. He received his BSc degree in Electronic Engineering in 2020 at Università di Modena e Reggio Emilia and he completed his MSc in Telecommunication Engineering in 2022 at Politecnico di Milano. 
\end{IEEEbiography}
\begin{IEEEbiography}[{\includegraphics[width=1in,height=1.25in,clip,keepaspectratio]{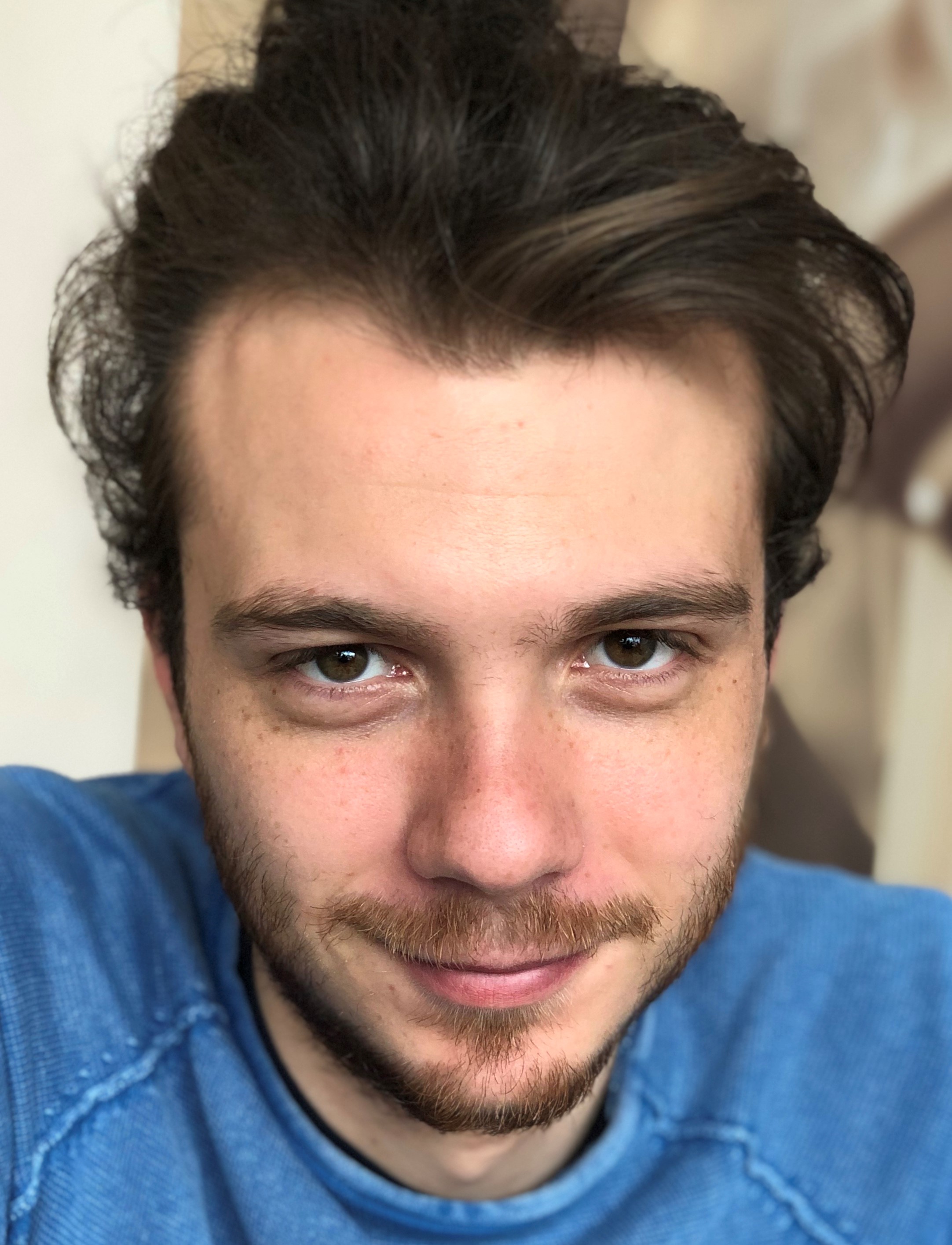}}]{Eugenio Moro} (Member, IEEE)  is currently an Assistant Professor at Politecnico di Milano, Milan, Italy. He received the M.Sc. and Ph.D. cum laude in 2019 and 2023, respectively. He was a visiting researcher in Nokia Bell Labs, Stuttgart, Germany and Northeastern University, Boston MA, USA. His research interests are wireless networks, with a focus on high-frequency radio access, wireless network programmability and optimization and smart propagation environments.
\end{IEEEbiography}
\begin{IEEEbiography}[{\includegraphics[width=1in,clip,keepaspectratio]{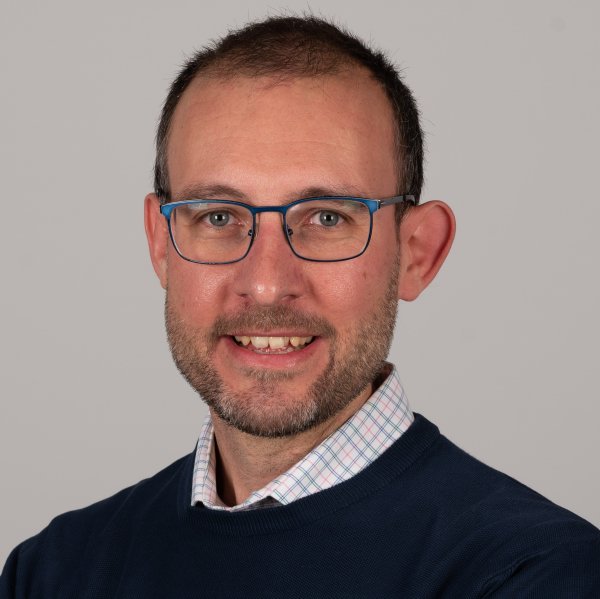}}]{Ilario Filippini} (Senior Member, IEEE) is an Associate Professor at the Dipartimento di Elettronica, Informazione e Bioingegneria of Politecnico di Milano. He received an M.S. in Telecommunication Engineering and a Ph.D. in Information Engineering from Politecnico di Milano in 2005 and 2009, respectively. From February 2008 to August 2008, he was visiting the Department of Electrical and Computer Engineering of The Ohio State University in Columbus (OH) working on routing in Cognitive Radio Networks.
He works on networking topics, his main research activities include radio resource management and optimization in wireless networks, programmable networks, and smart radio environments. He has co-authored more than 70 peer-reviewed conference and journal papers. He serves in the Technical Program Committee of major conferences in Networking and as an Editor of Computer Networks (Elsevier).
\end{IEEEbiography}
\begin{IEEEbiography}[{\includegraphics[width=1in,height=1.25in,clip,keepaspectratio]{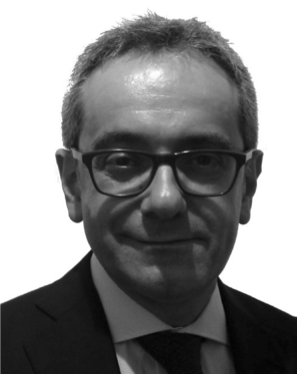}}] {Antonio Capone} (Fellow, IEEE) is currently the Dean of the School of Industrial and Information Engineering, Politecnico di Milano (Technical University of Milan). His main research activities include radio resource management in wireless networks, traffic management in software defined networks, network planning, and optimization. He is author of over 250 publications on these topics. He contributes to major international conferences on networking as a technical program committee member. He is an Editor of the IEEE
TRANSACTIONS ON MOBILE COMPUTING, Computer Networks, and Computer Communications and served as an Editor for the ACM/IEEE TRANSACTIONS ON NETWORKING from 2010 to 2014.
\end{IEEEbiography}
\begin{IEEEbiography}[{\includegraphics[width=1in,height=1.25in,clip,keepaspectratio]{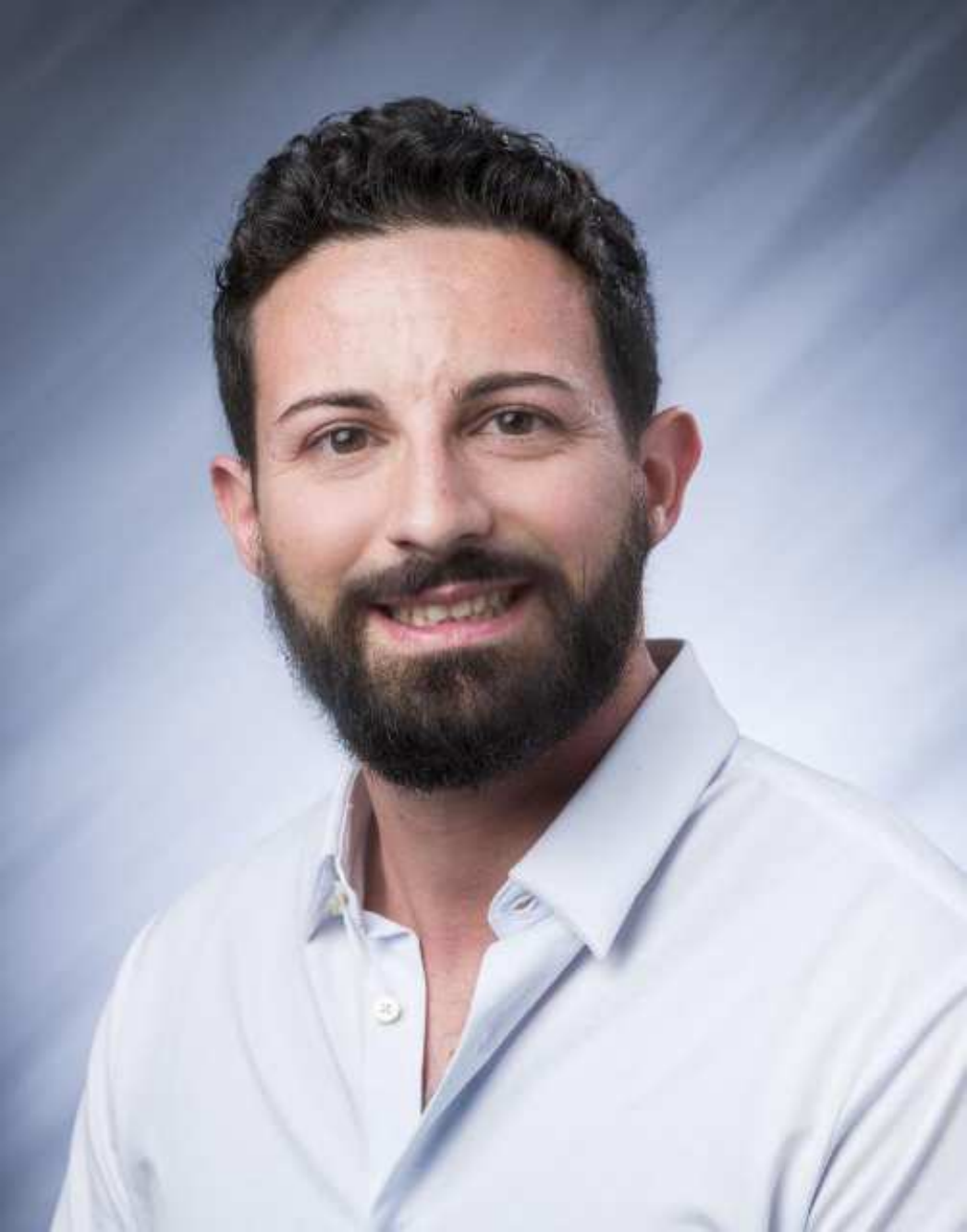}}]{Danilo De Donno} received the B.Sc. and M.Sc.
degrees in Telecommunication Engineering from Politecnico di Milano, Italy, in 2005 and 2008, respectively, and the Ph.D. degree in Information Engineering from the University of Salento, Lecce, Italy, in 2012. He was a Post-Doctoral Fellow with
the ElectroMagnetic Lab Lecce (EML2) of the University of Salento from 2012 to 2015 and a Post-Doc Researcher with the IMDEA Networks Institute, Madrid, Spain, from 2015 to 2017. In July 2017, he joined the Huawei Research Center in Milan, Italy, as a Wireless System Engineer. His main research focus is on mmWave technologies, systems and networks, beyond-5G RAN topologies, optimization of Smart Electromagnetic Environment. 
\end{IEEEbiography}
\end{document}